\documentclass[twocolumn]{aastex63}

\usepackage{enumerate}
\usepackage{multirow}
\usepackage{graphicx}
\usepackage{lineno}
\usepackage[normalem]{ulem}
\usepackage{xcolor}
\usepackage[colorlinks=true]{hyperref}
\usepackage{tabularx}
\usepackage{verbatim}
\usepackage{booktabs}
\usepackage{afterpage}
\hypersetup{linkcolor=blue,citecolor=blue,filecolor=cyan,urlcolor=blue}
\usepackage{amsmath}
\usepackage{epstopdf} 
\usepackage{kotex}
\usepackage{mathrsfs}

\newcommand*\numcircledmod[1]{\raisebox{.5pt}{\textcircled{\raisebox{-.9pt} {#1}}}}

\newcommand{\pgalf}{{\texttt{pGalF}}}
\newcommand{\psb}{{\texttt{PSB}}}
\newcommand{\ramses}{{\texttt{RAMSES}}}

\newcommand{\msun}{\mathrm M_{\odot}}
\newcommand{\msunh}{{h^{-1} {\mathrm M_{\odot}}}}
\newcommand{\kms}{{\mathrm km\,s^{-1}}}
\newcommand{\cmpch}{h^{-1} {\rm cMpc}}

\newcommand{\hr}{\texttt{HR5}}

\definecolor{red}{RGB}{250,0,0}
\definecolor{blue}{RGB}{0,0,255}
\definecolor{green}{RGB}{48,127,0}

\shorttitle{Missing LSBGs in the observed GSMF}
\shortauthors{Kim et al.}

\begin{document}

\title{Low-Surface-Brightness Galaxies are missing in the observed Stellar Mass Function}
\author[0000-0002-4391-2275]{Juhan Kim}
\affiliation{Center for Advanced Computation, Korea Institute for Advanced Study,  
85 Hoegiro, Dongdaemun-gu, Seoul 02455, Republic of Korea}
\author[0000-0002-6810-1778]{Jaehyun Lee}
\affiliation{Korea Institute for Advanced Study, 85 Hoegi-ro, Dongdaemun-gu, Seoul 02455, Republic of Korea}
\author{Clotilde Laigle}
\author[0000-0003-0225-6387]{Yohan Dubois}
\affiliation{CNRS and Sorbonne Universit\'e, UMR 7095, Institut d'Astrophysique de Paris, 98 bis, Boulevard Arago, F-75014 Paris, France}
\author{Yonghwi Kim}
\affiliation{Korea Institute for Advanced Study, 85 Hoegi-ro, Dongdaemun-gu, Seoul 02455, Republic of Korea}
\affiliation{Department of Astronomy, Yonsei University, 50 Yonsei-ro, Seodaemun-gu, Seoul 03722, Republic of Korea}
\author[0000-0001-9521-6397]{Changbom Park}
\affiliation{Korea Institute for Advanced Study, 85 Hoegi-ro, Dongdaemun-gu, Seoul 02455, Republic of Korea}
\author[0000-0003-0695-6735]{Christophe Pichon}
\affiliation{CNRS and Sorbonne Universit\'e, UMR 7095, Institut d'Astrophysique de Paris, 98 bis, Boulevard Arago, F-75014 Paris, France}
\affiliation{IPhT, DRF-INP, UMR 3680, CEA, L'Orme des Merisiers, B\^at 774, 91191 Gif-sur-Yvette, France}
\affiliation{Korea Institute for Advanced Study, 85 Hoegi-ro, Dongdaemun-gu, Seoul 02455, Republic of Korea}
\author[0000-0003-4446-3130]{Brad K. Gibson}
\author{C. Gareth Few}
\affiliation{E.A. Milne Centre for Astrophysics, University of Hull, Hull, HU6 7RX, United Kingdom}
\author[0000-0001-5135-1693]{Jihye Shin}
\affiliation{Korea Astronomy and Space Science Institute, 776 Daedeokdae-ro, Yuseong-gu, Daejeon 34055, Republic of Korea}
\author{Owain Snaith}
\affiliation{University of Exeter, School of Physics and Astronomy, Stocker Road, Exeter, EX4 4QL, UK}

%\author[0000-0003-0695-6735]{Christophe Pichon}
%\affiliation{CNRS and Sorbonne Universit\'e, UMR 7095, Institut d'Astrophysique de Paris, 98 bis, Boulevard Arago, F-75014 Paris, France}
%\affiliation{ IPhT, DRF-INP, UMR 3680, CEA, L'Orme des Merisiers, B\^at 774, 91191 Gif-sur-Yvette, France}
%\affiliation{Korea Institute for Advanced Study, 85 Hoegi-ro, Dongdaemun-gu, Seoul 02455, Republic of Korea}

\correspondingauthor{Jaehyun Lee}
\email{syncphy@gmail.com}

\begin{abstract}
We investigate the impact of the surface brightness (SB) limit on the galaxy stellar mass functions (GSMFs) using galaxy catalogs generated from the \texttt{Horizon Run 5} (\hr) simulation. We compare the stellar-to-halo-mass relation, GSMF, and size-stellar mass relation of the \hr\ galaxies with observational data and other cosmological simulations. The mean SB of simulated galaxies are computed using their effective radii, luminosities, and colors. To examine the cosmic SB dimming effect, we compute $k$--corrections from the spectral energy distributions of individual simulated galaxy at each redshift, apply the $k$--corrections to the galaxies, and conduct mock surveys based on the various SB limits. We find that the GSMFs are significantly affected by the SB limits at a low-mass end. This approach can ease the discrepancy between the GSMFs obtained from simulations and observations at $0.6\lesssim z\le 2$. We also find that a redshift survey with a SB selection limit of $\left<\mu_r\right>^e =$ 25 mag arcsec${}^{-2}$ will miss 20 \% of galaxies with $M_\star^g=10^{9}~{\rm M_\odot}$ at $z=0.625$. The missing fraction of low-surface-brightness galaxies increases to 35 \%, 55 \%, and 80 \% at $z=0.9$, 1.1, and 1.9, respectively, at the same SB limit.
\end{abstract}


%\maketitle

\section{\label{sec:in}Introduction}
The observed galaxy stellar mass function \citep[GSMF; for a non-complete list of references, see][]{bell+03,Baldry+08,weigel+16,Davidzon_2017,weaver22} has usually been regarded as an essential tool for understanding galaxy evolution and star formation history \citep{madau+14,katsianis+15, boco+21, Adams21, Puchwein+13, conroy+07, Grazian_2015}. The galaxy stellar mass is derived from the color-dependent mass-to-light ratio \citep{bell+01, madau+14} or the galaxy spectral energy distribution \citep[SED;][]{conroy+13, panter+07, bolzonella+10} where the galaxy history of star formation activities \citep{sidney+20, madau+14} are encoded. We are able to study the evolution of the baryonic content in galaxies using this encrypted global star formation history or the GSMF. 

In the cosmological context, galaxies are believed to form inside dark matter halos \citep[for a recent review on the galaxy \& halo relation, see][]{wechsler+18}. The GSMFs are accordingly expected to follow the halo mass functions (HMFs), but their shapes are quite different to each other. The HMF rises more steeply in the low-mass tail (see \citealt{Kim+15} for a non-complete list of various fitting functions) while the GSMF has a much shallower (power-law) tail below the knee of the Schechter function \citep[][]{Davidzon_2017,Adams21,Song_2016, McLeod+21, weigel+16}. On the low-mass scale, this discrepancy is mainly attributed to stellar feedback \citep[to name a few]{Benson+03,Baldry+08,Silk+12, Puchwein+13}. Supernova (SN) explosions and stellar winds heat and eject the interstellar medium from galaxies in low mass halos, and accordingly suppressing the growth of galaxy~\citep[e.g.][]{Silk+12}. 

Motivated by this, some efforts have been make to reproduce the observed GSMFs by implementing stellar feedback into gravito-hydrodynamics simulation~\citep{vogelsberger+13,dubois+14,katsianis+15,schaye+15,dubois16,pillepich18}. The \texttt{Illustris TNG} simulations are updated from the Illustris simulations~\citep{Vogelsberger+14} to better fit the low-$z$ GSMFs. They reduced the star formation in low-mass galaxies by adopting a redshift-dependent wind velocity floor in the mass loading factor \citep[for details, see][]{Springel18,pillepich18,Naiman18,Marinacci18,nelson18}. The \texttt{EAGLE} simulation focused on stochastic thermal feedback to prevent the overcooling problem, without turning-off the radiative cooling \citep[for details, see][]{schaye+15,crain15}. They showed that \texttt{EAGLE} could well reproduce the GSMF for small-mass galaxies. The \texttt{Horizon-AGN} \citep[hereafter \texttt{H-AGN,}][]{dubois+14} simulation adopts a dual mode of SN feedback (kinetic and thermal). %The relative contributions of the two modes are tuned by a free-parameter.

The low-mass end slope of a GSMF has been debated due to incomplete observations for low-surface-brightness galaxies~\citep[LSBGs;][]{Baldry+08, Tang+21, martin19, valss-gabaud+17, greene+22}. A well known example of this issue in low redshift surveys (typically with the surface brightness of $\left<\mu_R\right>^e \ge 21.82$ mag arcsec${}^{-2}$) is given in the figure 10 of \cite{geller+12}. It shows that the LSBGs dominate the faint-end regime of the galaxy luminosity functions (LFs). Therefore, a somewhat shallow surface brightness (SB) cut probably misses the majority of LSBGs at the low brightness end of the galaxy LFs.

The pencil-beam surveys have achieved deep SB limits above the background fluctuations, at the cost of the area of fields~\citep[see][]{lee+12}. For example, \cite{reis+20} identified high-$z$ galaxies from the HST CANDELS archive image data with the SB limits of 28.3 mag arcsec$^{-2}$ $\le \left<\mu_{\rm F160W}\right>\le 28.6$ mag arcsec${}^{-2}$ corresponding to the $3\sigma$ limiting magnitude in the COSMOS, GOODS-N, UDS, and EGS fields for an aperture size of $4{\arcsec}$ (main) and $1{\arcsec}$ (small galaxies). \cite{Grazian_2015} adopt 27 mag arcsec$^{-2} \le \left<\mu_{\rm F160W}\right>\le 28.5$ mag arcsec${}^{-2}$ at 1$\sigma$ limit in GOODS-S for galaxies at $3.5\le z\le 7.5$ with aperture size $2\times$ the full width at half maximum (FWHM). \cite{Adams21} derived the GSMFs from the HSC (Hyper Suprime-Cam) data of the COSMOS \& XMM-LSS fields with the SB limit of 25.9 mag arcsec${}^{-2}$  $\le \left<\mu_r\right>^0\le$ 26.7 mag arcsec${}^{-2}$ at 5$\sigma$ with the aperture of $2{\arcsec}$ diameter.  \cite{tomczak+14} measured the GSMF from NIR imaging of the CDFS, COSMOS, and UDS fields with Magellan Baade telescope, which reaches the $5\sigma$ depths of $24.6< \left<\mu_{K_s}\right><25.2$ in a circular diameter of $0\arcsec.6$.   

On the other hand, wide-field surveys impose relatively shallow SB limits to the source detection criteria, to achieve high statistical significance in galaxy sampling. To list a few, the SDSS has the SB selection criterion of $\left<\mu_r\right>^e \le 23$ mag arcsec${}^{-2}$ \citep{blanton+05}. The SB selection criteria are $\left<\mu_r\right>\le 23$ mag arcsec${}^{-2}$ for the LCRS \citep{cross+01}, $\left<\mu_{bJ}\right>^e\le 24.7$ mag arcsec${}^{-2}$ for 2dFGRS \& APM \citep{cross+01}, $\left<\mu_r\right>^{\rm  petro}\le 21.3$ mag arcsec${}^{-2}$ for HectoMAP \citep{sohn+18}, and $\left<\mu_R\right>^e\le 21.82$ mag arcsec${}^{-2}$ for SHELS${}_{0.1}$ \citep{geller+12}. The wide field surveys enable us to obtain accurate GSMFs in the local universe, but they are still incomplete to detect LSBGs due to their shallow detection limits. Given that the GSMF is the key statistics in calibrating physical ingredients of galaxy formation models, it necessitates to evaluate the missing LSBG populations in the observed GSMFs. This study aims at quantifying the fraction of missing LSBGs as a function of the SB detection limit using the \texttt{Horizon Run 5}~\citep[hereafter \hr,][]{lee21,park22} simulation.

This paper is organized as follows. In Section \ref{sec:1}, we describe the \hr\ simulation and its output galaxy catalog. The galaxy properties of \hr\ are statistically analyzed in Section \ref{sec:3}, and the high-redshift and the low-redshift GSMFs are fully addressed in Sections \ref{sec:4} and \ref{sec:5}, respectively. We close this study with discussions in Section \ref{sec:dis}. Additionally we describe the galaxy finding method used for \hr\ in \ref{sec:finder}. Also, we provide auxiliary calculations on the magnitude transform (\ref{sec:diffmag}), source detection criteria (\ref{sec:source}), and the dependence of GSMFs on the star formation efficiency (\ref{sec:sfecomp}) and the definition of the effective radius (\ref{sec:sizediff}).

\section{Simulation}
\label{sec:1}

\subsection{Horizon Run 5}

\hr\ is a cosmological hydrodynamic simulation aiming at studying galaxy formation and evolution from high- to intermediate-redshift (down to $z=0.625$) in a cubic volume with a side length of $L_\mathrm{box} = 1049 ~\mathrm{cMpc}$. A zoomed region is set to have a cuboid geometry of $L_\mathrm{(x,y,z)}^\mathrm{zoom} = (1049, 119, 127 )$ cMpc crossing the central region of the simulation box. Detailed information of the adopted cosmology and simulation setup can be found in \cite{lee21}. We identify halos and galaxies from the entire snapshots of \hr\ using the Physically Self-Bound (PSB)-based galaxy finder (\pgalf). A detailed description of \pgalf\ are given in \ref{sec:finder}.

\subsection{Star Formation and Feedback}
We summarize the sub-grid physics associated with star formation and stellar feedback adopted in \hr\ here. Star formation rates are computed based on a Schmidt law~\citep{schmidt59}:
\begin{equation}
	\frac{d\rho_{\star}}{dt}=\epsilon_{\star}\frac{\rho_{\rm gas}}{t_{\rm ff}},
\end{equation}
where $\rho_{\rm gas}$ is the gas density in a cell, $\epsilon_{\star}$ is the star formation efficiency per freefall time, and $t_{\rm ff}=\sqrt{3\pi/32G\rho_{\rm gas}}$ is the freefall time of a gas cell, where $G$ is the gravitational constant. We adopt a constant star formation efficiency of $\epsilon_{\star}=2\%$ to reproduce the global SFR evolution. %For detailed conditions and numerical parameters, see \cite{lee21}. 

The version of \ramses\ adopted for \hr\ has two modes of AGN feedback switched by the Eddington ratio~\citep{dubois+12}, 
\begin{equation}
	\chi \equiv \left( { \dot{ M}_{\rm BH} \over\dot{M}_\mathrm{Edd} }\right),
\end{equation}
where $\dot{ M}_{\rm BH}$ is the growth rate of black hole mass and $\dot{ M}_\mathrm{Edd}$ is the Eddington limit. If $\chi \leq \chi_{\rm c, jet}=0.01$, the radio (dual jet--heating) mode is turned on, and otherwise, the quasar (thermal) mode operates. The total energy released in the thermal form is $\dot{E}_\mathrm{BH,h} = \epsilon_{\rm r} \epsilon_{\rm f,h} \dot{M}_{\rm BHL} c^2$ where $\epsilon_{\rm r}$ is the spin-dependent radiative efficiency, $\epsilon_{\rm f,h}$ is the coupling efficiency of the thermal feedback, $\dot{M}_{\rm BHL}$ is the Bondi-Hoyle-Lyttleton accretion rate, and $c$ is the speed of light. A jet mode releases the amount of energy of $\dot{E}_{\rm BH,j} = \epsilon_{\rm f,j}\dot{M}_{\rm BHL} c^2$, where $\epsilon_{\rm f,j}$ is the spin-dependent coupling efficiency of the jet mode~\citep{mckinney12}. Therefore, AGN feedback is controlled by the two free parameters, $\chi_{\rm c, jet}$ and $\epsilon_{\rm f,h}$. We adopt $\epsilon_{\rm f,h}=0.15$ to reproduce the observed $M_{\rm BH}$--$M_\star$ relation at $z=0$ \citep{booth+09, dubois+12} and $\chi_{\rm c, jet}=0.01$~\citep{merloni+08}. The gas accretion rate is capped by the Eddington limit in \hr. For more details of AGN feedback in \hr, see \citet{dubois14a} and \citet{dubois20}.

In \hr, we assume stellar feedback operated by the winds from asymptotic giant branch (AGB) stars and Type Ia and Type II supernovae (SNe). The amount of energy released from an SN is set to $2\times 10^{51}$~erg. When stellar particles are sufficiently young, we deposit 30\% of the SNII energy as kinetic energy and the rest as thermal. Old stellar populations release their energy in the SNIa and AGB wind modes, deposit the energy in its thermal form, and eject their mass to the nearest grid cells. The initial mass function (IMF) of stellar populations is one of primary parameters governing stellar feedback. We adopt a Chabrier IMF~\citep{chabrier03} for all stellar populations in \hr.

\subsection{Size \& Brightness of Simulated Galaxies}
We define the galaxy size as the half mass radius of stellar components projected on an $XY$ plane of the simulation coordinate. We assume the density peak of stellar mass distribution as the center of a galaxy. The rest-frame SEDs of galaxies are modeled using the masses, ages and metallicities of stellar particles based on the \texttt{E-MILES} single stellar population SEDs~\citep{vazdekis12,vazdekis16,ricciardelli12}. We assume a Chabrier IMF in this calculation, for consistency with the stellar population model adopted in \hr. We also compute the luminosities of galaxies using the photometric predictions of \texttt{E-MILES} in the \texttt{Johnson} and \texttt{SDSS} filter systems.

\section{Distributions of Galaxy Properties}
\label{sec:3}

\subsection{Aperture Correction}
Numerical simulations have suffered from the excess of BCG-scale galaxies, compared to the empirical GSMFs, and thus the idea of contained mass has been proposed to reconcile the discrepancy~(\citealt{schaye+15,MaCarthy+17,Adams21,Tang+21}, and for an extensive discussion on the choice of the aperture size, see \citealt{donnari+19}). The BCG of a cluster can have stellar mass overestimated due to its far-extended stellar envelope tangled with the intra-cluster light \citep{zwicky51,gonzalez+07,guennou+12,mihos19,yoo+21,montes22}. For instance, \citet{pillepich18} demonstrated that the galaxies of total stellar mass of $M\sim10^{12}\,\msun$ have half their mass outside 30 pkpc. Observationally, however, the galaxy stellar mass is generally measured in the Petrosian aperture or by assuming a S\'ersic-model profile \citep{Graham_2005}, both of which may neglect the extended stellar components of the BCGs. Following the idea of the aperture correction, we also adopt the 3-D aperture of a 30 pkpc radius, to alleviate the mass overestimation of BCG-scale galaxies.
\subsection{Relation between Stellar Brightness and Mass }
The global star formation rate (SFR) of the universe peaks at $z\sim$ 2--3 and declines after the cosmic noon~\citep{Hopkins04,behroozi13,madau+14}, resulting in the increase of the mass-to-light ratios of galaxies \citep{vandeven+03,vanderwel+05}. 
\begin{figure}
\centering 
\includegraphics[width=\linewidth]{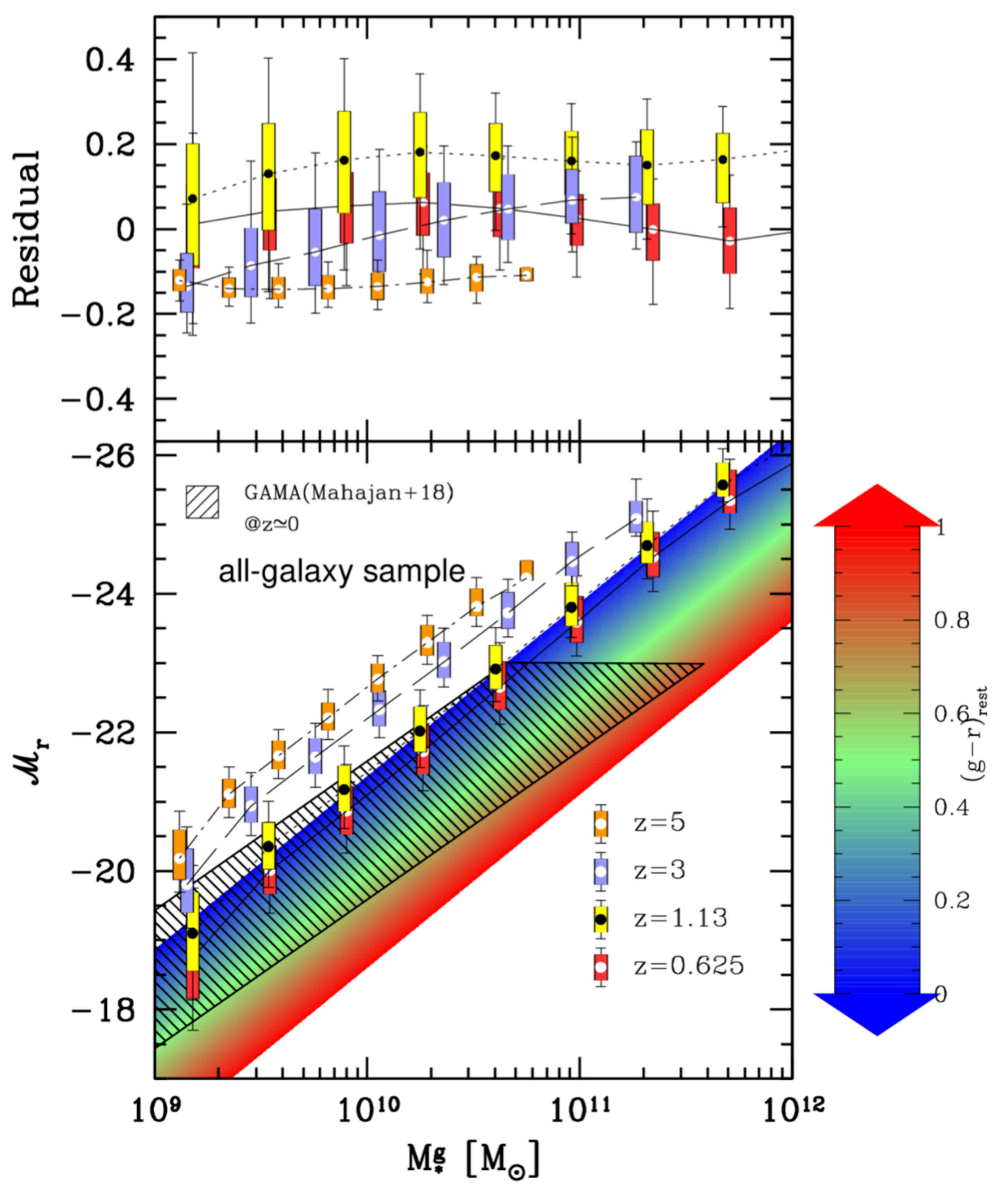}
\caption{({\it bottom}): Redshift evolution of the galaxy stellar-mass function with respect to the absolute magnitude in the rest-frame $r$ band. For comparison, we overlap the observed relation for blue spheroids from the GAMA survey \citep[the hatched area,][]{mahajan+18} with our fitting model simply extrapolated to $z=0$ (the colored region). ({\it top}): Fitting residuals. The filled circle is the median in each stellar-mass bin and the colored box is the 1--$\sigma$ distribution. The error bar marks the 2--$\sigma$ scattering.}
\label{fig:mag}
\end{figure}

\cite{bell+03} presented the relation between the galaxy stellar mass $M_\star^g$ and the $r$-band absolute magnitude ($\mathcal{M}_r$) as a function of redshift and rest-frame galaxy color ($g-r$) as
\begin{eqnarray}\nonumber
	\log_{10}\!\left({ M_\star^g \over M_{\rm \odot}}\right) &=& 1.097(g-r) \!-\! 0.406\! -\!0.4(\mathcal{M}_r\!-\!4.67) \\
	&& + f_{\rm Bell}(z),
	\label{eq:bernardi10}
\end{eqnarray}
where $f_{\rm Bell}(z)=-0.19z$ reflects the redshift evolution in the 2MASS/SDSS galaxies. However, this relation is based on the sample of the local universe and only applicable to low redshift galaxies. Hence, instead of $f_{\rm Bell}$, we propose a new fitting model of 
\begin{equation}
f_{\rm HR5}(z)=- \left\{0.02+4.96(g-r)\right\} \log_{10}(1+z),
\end{equation}
which is formulated to capture the co-evolution of redshift and galaxy color at $0.625 \le z \le 5$. 

Figure~\ref{fig:mag} shows the evolution of the mass-to-brightness relation at given galaxy stellar mass as a function of redshift. In this plot, the r-band magnitude $\mathcal{M}_r$ is given at fixed $M_\star^g$ because galaxy stellar mass is intrinsic in simulations while the brightness is derived based on assumed stellar models. Throughout this paper, we maintain this convention.

At $z=0.625$, galaxies tend to be fainter than their higher-redshift counterparts at a fixed mass due to aging of the stellar populations. The aging effect is taken into consideration by the correction term ($f_{\rm Bell}$ or $f_{\rm HR5}$). In the top panel of the figure, we show the residuals of the fit. Here the residual absolute magnitude ($\mathcal{R}$) is the difference of the absolute magnitude between the simulated galaxy and the modeled one and is written as
\begin{equation}
	\mathcal{R}(M^g_\star) \equiv \mathcal{M}_r({\rm sim}) - \mathcal{M}_r({\rm model}),
\end{equation} 
for a given galaxy stellar mass. The majority of the 1--$\sigma$ scatterings are within the 0.2 magnitude error.

\subsection{Environmental Dependence of Galaxy Brightness}
Now, we present the environmental effects on galaxy absolute magnitude in the SDSS $r$-band. From the simulation catalog of \hr, we divide entire galaxy sample ($G_\texttt{all}$) into the satellite ($G_\texttt{sat}$), central ($G_\texttt{cen}$), and isolated ($G_\texttt{iso}$) samples. The most massive galaxy in each halo is classified as centrals, while the satellite sample contains the rest, less-massive galaxies. The isolated galaxies are those with no companion in their FoF halos.

\begin{figure}
\centering 
\includegraphics[width=\linewidth]{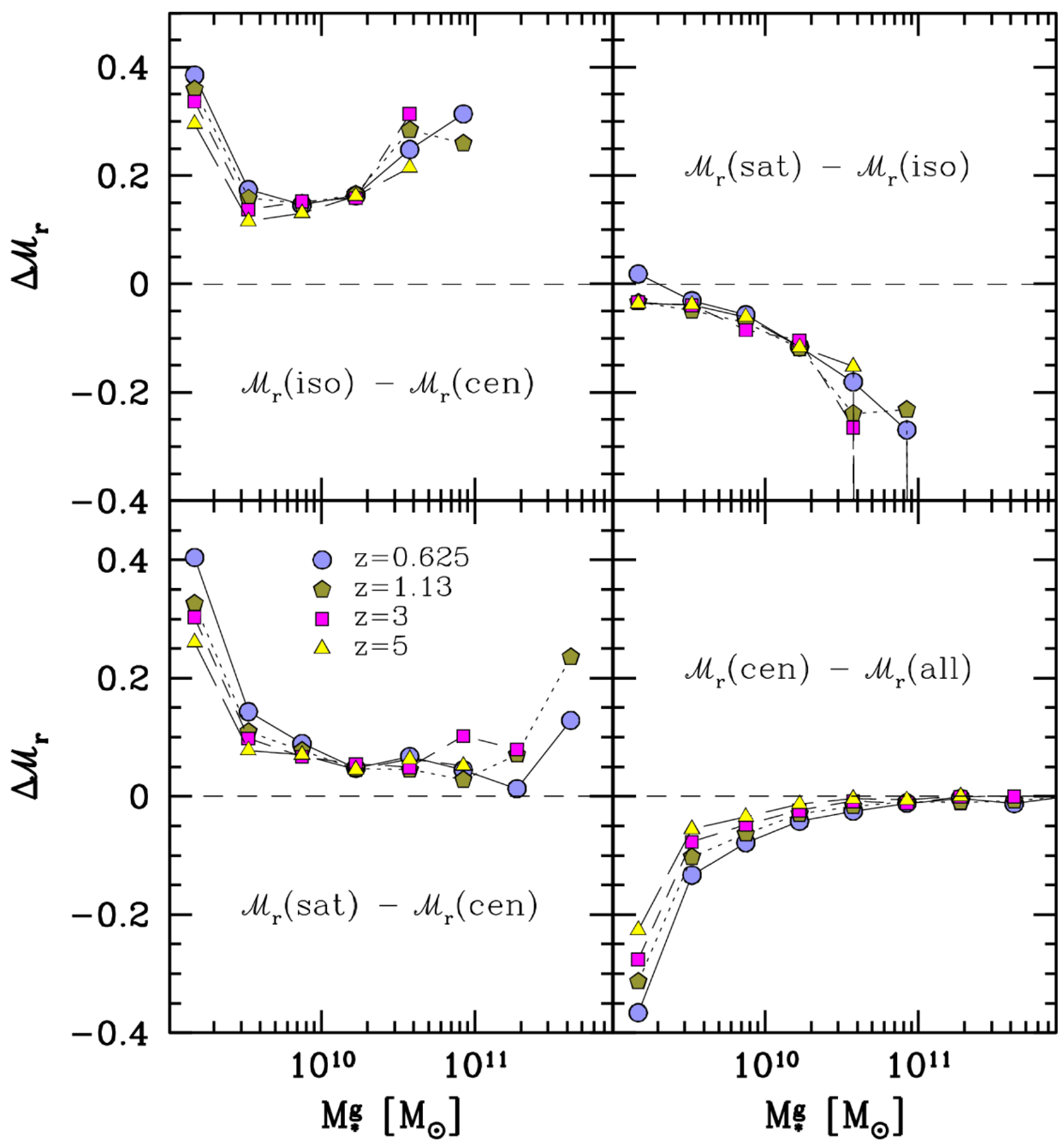}
\caption{Relation between environments and galaxy brightness as a function of galaxy stellar mass. The differences of average absolute magnitudes are shown between $G_\texttt{cen}$ \& $G_\texttt{all}$, $G_\texttt{sat}$ \& $G_\texttt{cen}$, $G_\texttt{cen}$ \& $G_\texttt{iso}$, and $G_\texttt{sat}$ \& $G_\texttt{sio}$ samples clockwise from the bottom-right panel, respectively.}
\label{fig:cen_mag}
\end{figure}

In the bottom right panel of Figure~\ref{fig:cen_mag}, $G_\texttt{cen}$ tend to be brighter than the average obtained from the $G_\texttt{all}$ sample, while this is less significant for more massive galaxies. Redshift-evolution of $\Delta \mathcal{M}_{\rm r}$ is apparent at the low-mass end due to the decline of SFRs in small satellite galaxies and the late formation of small central galaxies. The bottom left panel shows this more clearly. Satellite galaxies are generally fainter than the central galaxies of the same mass, and this tendency is more pronounced at a low-mass end. 

The two top panels demonstrate that the isolated galaxies ($G_\texttt{iso}$) are fainter than both the central and satellite galaxies. Given that a younger stellar population is brighter at a fixed mass, these results indicate that star formation activities are lowered in the isolated galaxies earlier than any other galaxy groups in \hr. The massive satellite galaxies are more likely to have been the centrals of groups or clusters until recently because their merging timescales are shorter than those of smaller ones~\citep[e.g.,][]{binney08,boylan-kolchin08}. This is consistent with the small magnitude difference between the centrals and satellites at the massive end. On the other hand, small satellites are not only vulnerable to environmental effects~\citep{samuel22} but also likely to have experienced environmental effects in their hosts for a long period of time due to long merging timescales~\citep{lee18}.
%Overall, we find that central galaxies are brighter than average, while satellite galaxies tend to have a suppressed SFR, while isolated galaxies show the least star formation activities at $0.625\le z\le 5$.

\begin{figure}
\centering 
\includegraphics[width=\linewidth]{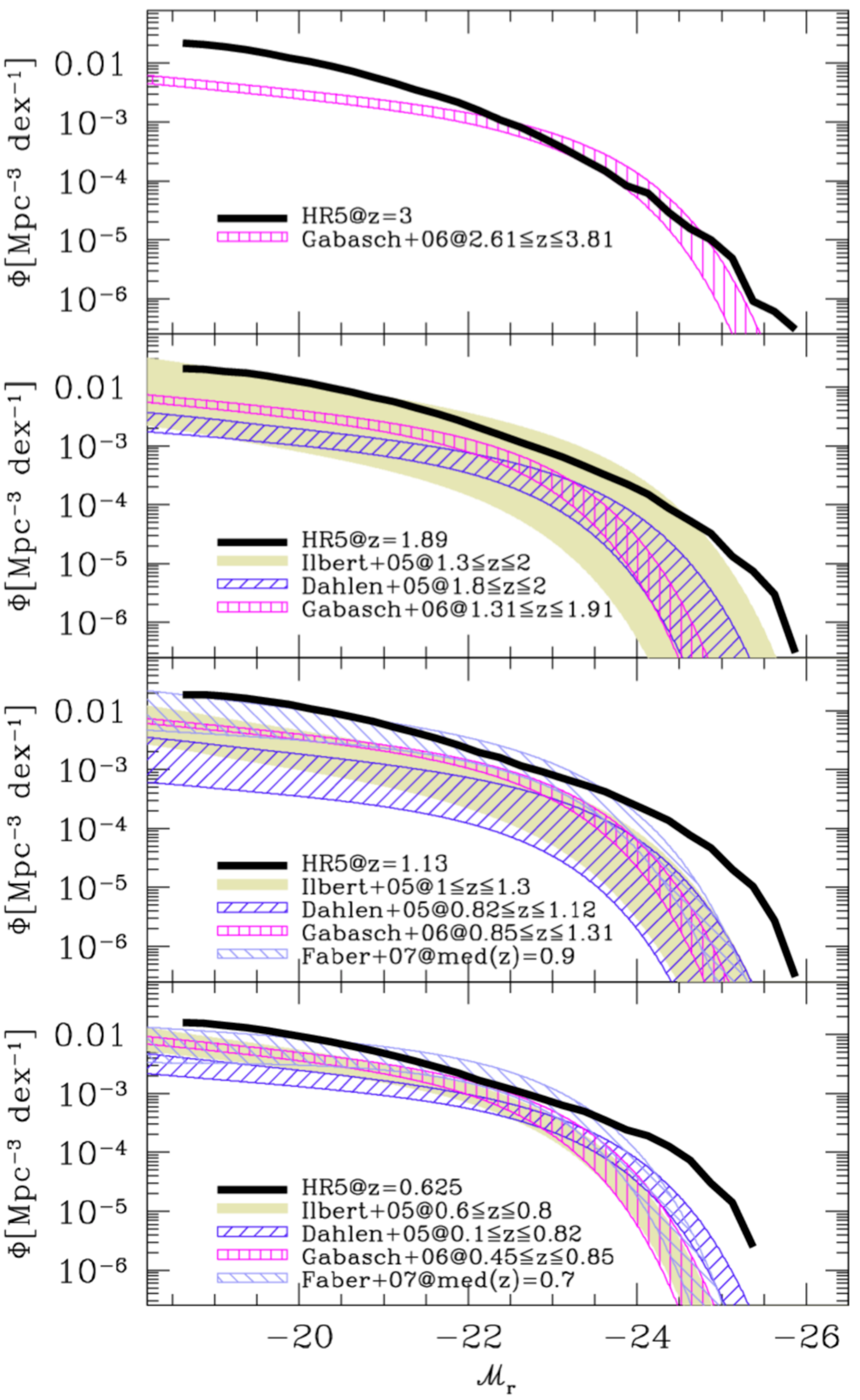}
\caption{Galaxy luminosity functions at $z=0.625$, 1.13, 1.89, and 3 (from the bottom panel). Thick solid lines are the mock galaxy luminosity functions from \hr. The hatched or shaded areas are observed Schechter functions (\citealt{ilbert05,dahlen05,gabasch06,faber07}).}
\label{fig:lf}
\end{figure} 

\subsection{Galaxy Luminosity Functions}
\cite{ilbert05} and \cite{dahlen05} derived galaxy LFs at $z\leq 2$ from the VIMOS-VLT Deep Survey, the HST ACS and GOODS South in the rest frame of R-band of the Cousins filter system. While the R-band has the similar wavelength coverage to that of the SDSS $r$-band, we have applied the $\chi^2$ fit to obtain the $r$-band luminosity functions (for the fitting form, see \ref{sec:diffmag}). We also add the LFs obtained from the FORS Deep Field data observed with the ESO VLT by \citet{gabasch06}, who derived the characteristic parameters of the Schechter function in the rest-frame $r$-band from the observations. Since the characteristic absolute magnitude of the Schechter function in \cite{faber07} are given in the $B_{\rm Johnson}$ band, we convert the SDSS $r$-band magnitudes of the \hr\ galaxies to the the $B_{\rm Johnson}$ band magnitudes by adopting the band transforms of $B_{\rm Johnson}= g+0.115+0.370\times 0.82$ and calculate $\mathcal{M}_r^\star$ by assuming the average color of $(g-r)=0.82$ (for details, see \citealt{faber07}).

We summarize the comparison between \hr\ and the observations in Figure \ref{fig:lf}, in which the LFs of \hr\ in the rest-frame $r$-band are illustrated in the thick black solid lines. As mentioned above, we apply the 30 pkpc aperture cut that is expected to reduce the total brightness of the BCGs substantially. In spite of the aperture cut, we still have substantial excess of the BCG-scale galaxies in \hr\ at $z\le 1.13$. Another notably feature seen in Figure~\ref{fig:lf} is that the excess of the simulated LFs to the observations becomes larger with increasing $\mathcal{M}_{\rm r}$ below the knee of the Schechter functions at all the redshifts. This is also seen in the galaxy LFs or GSMFs of \texttt{H-AGN}~\citep[see][]{kaviraj17}. \citet{kaviraj+17} speculated that the excess of low-mass galaxies is attributed to insufficient stellar feedback in \texttt{H-AGN}. This systematic excess on both mass or luminosity scales will be addressed quantitatively in the following sections.

\subsection{Stellar Mass versus Halo Mass Relation}
The stellar-to-halo-mass (SHM) relation is one of key properties that cosmological hydrodynamical simulations should be able to reproduce. The SHM relation connects galaxy observations to $N$-body simulations, which enables one to pin down the cosmological models and to study the coupled evolution of halos and galaxies. 

\begin{figure}
\centering 
\includegraphics[width=\linewidth]{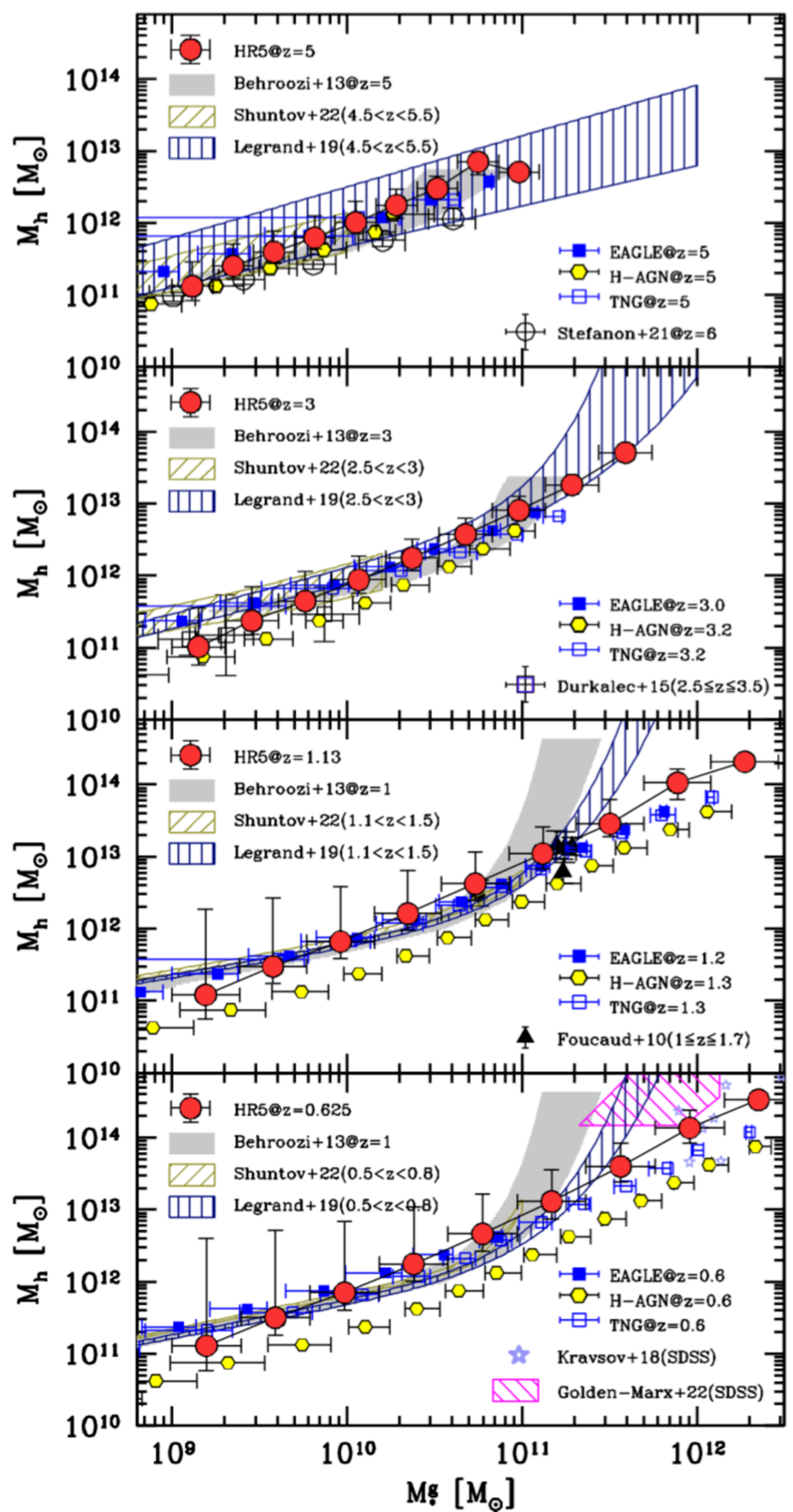}
\caption{Relations between $M_h$ and $M_*^g$. Symbols with error bars are the medians and 1--$\sigma$ scatters in each galaxy mass bin of the \hr (red filled circles), \texttt{EAGLE} (blue filled squared), \texttt{TNG100} (blue open squares), and \texttt{H-AGN}(yellow filled hexagons) at $z\sim$ 0.6, 1.3, 3, and 5 (from bottom to top). In this plot, we do not apply any aperture corrections to \hr, for a fair comparison with other simulations. Other simulation results are those provided by \cite{shuntov+22}. {\it Observations}: \citet{Golden_Marx_2022} (pink hatched, $z\sim0$), \citet{behroozi13} (grey shades), \citet{foucaud10} (black triangles, $z~1.3$), \citet{durkalec+15} (open navy squares, $z\sim3$ ), \citet{kravtsov+18}(open stars, $z\sim0$), \citet{shuntov+22}(green hatched), \citet{legrand+19} (vertically hatched), and \citet{stefanon21}(open circles, $z=6$).}
\label{fig:msmh}
\end{figure}

\begin{figure}
\centering 
\includegraphics[width=\linewidth]{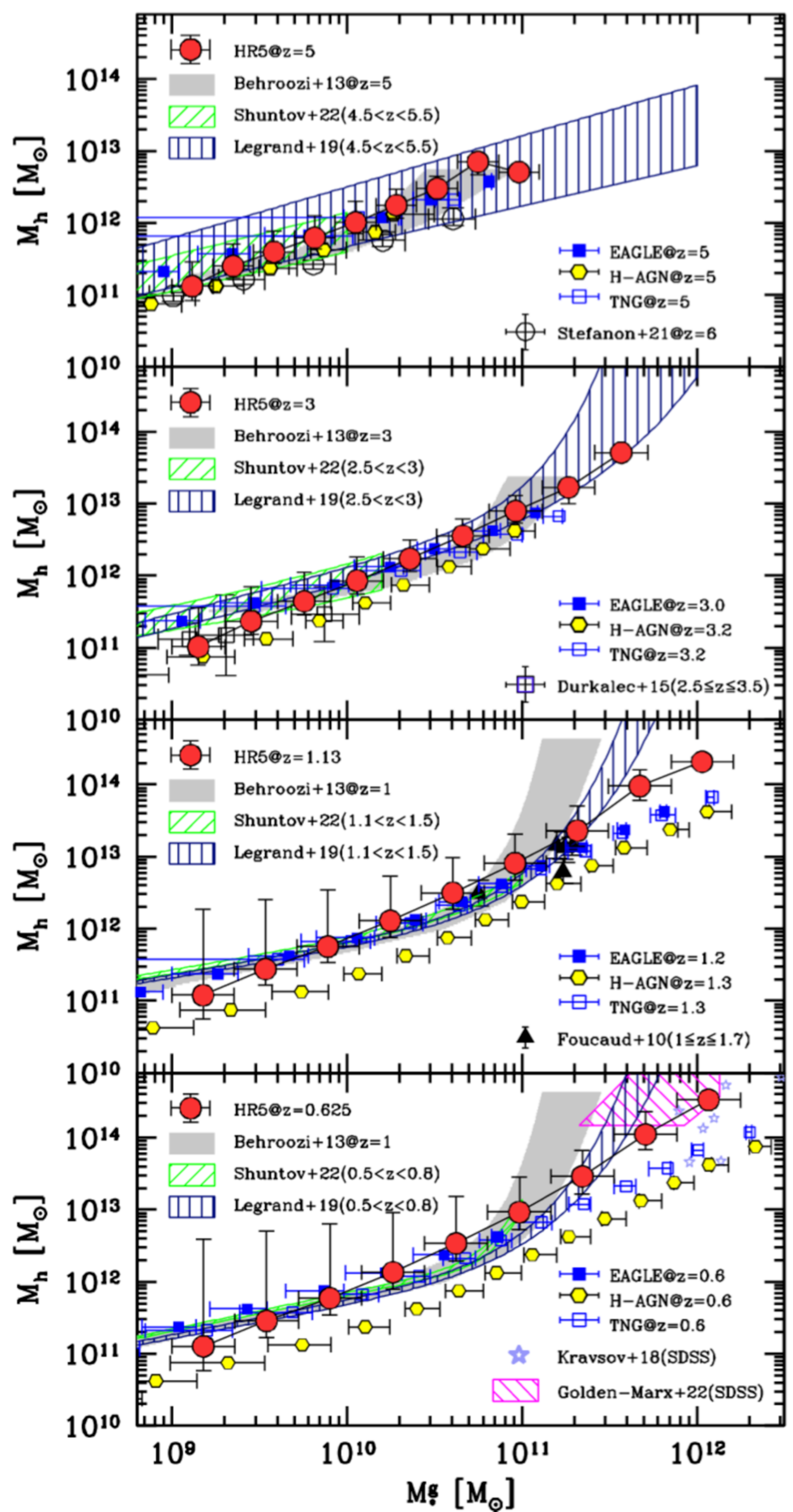}
\caption{Same as Figure~\ref{fig:msmh} but with the 30 pkpc aperture correction only to \hr.}
\label{fig:msmh_30kpc}
\end{figure}

Figure \ref{fig:msmh} shows the SHM relation of four different simulations (filled circles with error bars) and several observations at $z\sim0$, 0.6, 1.3, 3, and 5. In this plot, no aperture cut is applied to \hr, for a fair comparison with other simulations. The results with an aperture cut is presented in Figure~\ref{fig:msmh_30kpc}. At $z=3$ and 5, the relations obtained from \hr\ well follow the observations, while at $z\sim1.3$ and 0.6, \hr\ overproduces the stellar mass on the BCG scale ($M\ge 3\times 10^{11} {\rm M_\odot}$ and see \citealt{bellstedt+16} for reference), compared to \cite{behroozi13} (the grey shaded regions). %However other observations \citep{kravtsov+18,Golden_Marx_2022,shuntov+22} seem to be comparable to the \hr\ results. 
\hr\ seems to be comparable with the SHM relations of \cite{kravtsov+18} (open blue stars) and \cite{Golden_Marx_2022} (pink hatched) at the massive end that are based on the local SDSS BCG catalogs ($0\le z \le 0.15$). The redshift evolution below $z\sim 1$ is known to be insignificant in the SHM relation~\citep{behroozi13,legrand+19,shuntov+22}, and thus the comparison between \hr\ at $z=0.625$ and the local observations is viable to some degree.

We also overplot the SHMs of other simulations, \texttt{EAGLE} \citep{schaye+15}, \texttt{H-AGN} \citep{dubois+14}, and \texttt{TNG100} \citep{pillepich18}. At all the redshifts investigated in this study, \texttt{EAGLE} and \texttt{TNG100} show the SHM distributions well agreeing with the empirical data, particularly in $M_\star^g\lesssim10^{11}\,\msun$.% because their parameters were extensively investigated to reproduce the empirical SHM at low redshift~\citep[see section 3 in][]{pillepich+18}. 
In \texttt{H-AGN}, galaxies are indeed by far too massive at fixed halo mass compared to other simulations and observations, especially at lower $z$. As mentioned above, this may be due to insufficient SN feedback (and consequently higher global star formation history) at high redshift~\citep{kaviraj+17}. On the other hand, all the simulations produce more stellar mass than the empirical relation at high halo-mass end at which AGN feedback dominantly regulates star formation; the slopes of SHM relation is not reproduced at the massive end at low $z$. In Figure \ref{fig:msmh_30kpc} we show the results of the 30 pkpc aperture correction applied only to \hr, where the discrepancy between simulations and observations at high-mass end is to some degree relieved. However, the slope still does not seem to be consistent with the empirical results.

\subsection{Effective Sizes of Galaxies}

\begin{figure}
\centering 
\includegraphics[width=\linewidth]{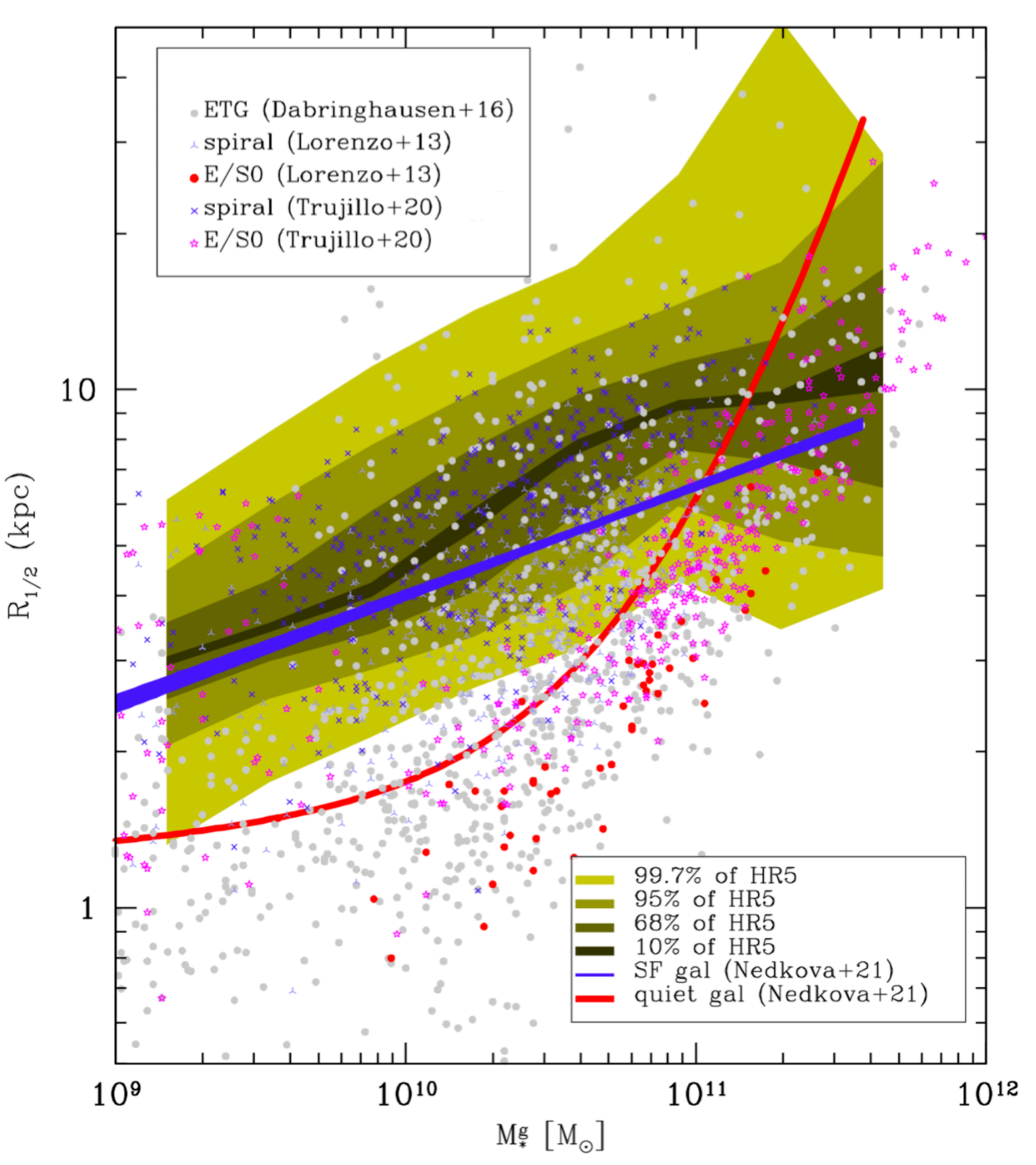}
\caption{Comparison of galaxy size distributions between observations and \hr\ at $z=0.625$. The background contour regions are the galaxy distributions of \hr\ in this plane. From the inner most contours, we show distribution of 10 \%, 68\% (1--$\sigma$), 95\% (2--$\sigma$), and 99.7\% (3--$\sigma$) around the median effective radius. {\it Observations}: \citet[lines]{Nedkova+21}, \citet{Trujillo+20}, \citet{lorenzo+13}, and \citet{dabringhausen+16}. }
\label{fig:galsize}
\end{figure} 

We look into the size distribution of \hr\ galaxies compared with observations in Figure \ref{fig:galsize}, where observations are plotted with various symbols. The background colored contour regions are obtained from the \hr\ galaxies at $z=0.625$. The lower outline of the size distributions of the \hr\ galaxies are above 1 pkpc, which is the limit imposed by the simulation resolution. This means that the 1 pkpc resolution is probably insufficient to properly simulate the early-type galaxies with mass below $M_\star^g\sim 10^{11}~{\rm M_\odot}$ (small red and gray circles in the figure) at $z=0.625$, suggesting the necessity of higher-resolution simulations.

\begin{figure}
\centering 
\includegraphics[width=\linewidth]{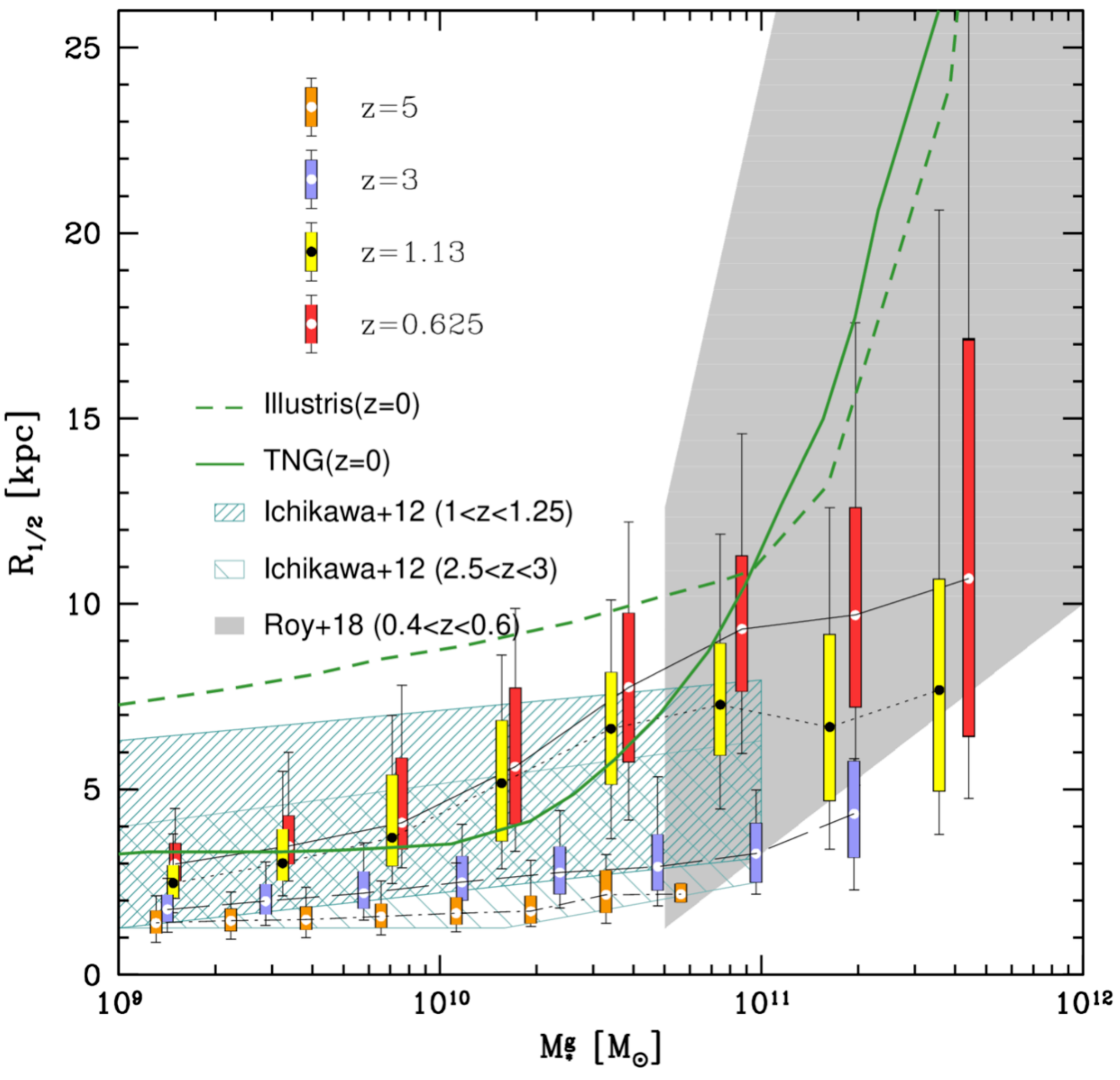}
\caption{Galaxy mass-size relation as a function of redshift. The distribution of galactic size for each stellar mass bin is shown with a filled circle (median), a colored box (1--$\sigma$), and an error bar (2--$\sigma$). The gray region is from the observations of \cite{roy+18} at $0.4<z<0.6$. and the hatch areas are obtained from  the star-forming galaxy samples in the GOODS-north region by \cite{ichikawa+12}. For simulation comparisons, we add the \texttt{Illustris} (thick-dashed) and \texttt{TNG100} (thick-solid line) results obtained from \cite{pillepich18}.}
\label{fig:psize}
\end{figure}

Figure~\ref{fig:psize} shows the time evolution of galaxy size $R_{1/2}$ as a function of stellar mass. At high redshift, galaxies are compact and the stellar-mass dependency is low, compared to those at lower redshift. However, as time goes by, the correlation between galaxy size and stellar mass becomes stronger~\citep[see also][]{dubois16}. The scatter of the galaxy-size distribution increases with increasing mass and decreasing redshift.

\subsection{A Model for k-Corrections}
Observers can obtain the rest-frame magnitude of a target object by applying the $k$-correction to the observed magnitude~\citep{blanton+05}. In the analysis of simulated galaxies, conversely, it is required to apply the $k$-correction in the opposite direction, to convert the rest-frame magnitude to the observed magnitude at a given redshift. In this study, we take the second approach to simulate the magnitude in observed frames.

\begin{figure}
\centering 
\includegraphics[width=\linewidth]{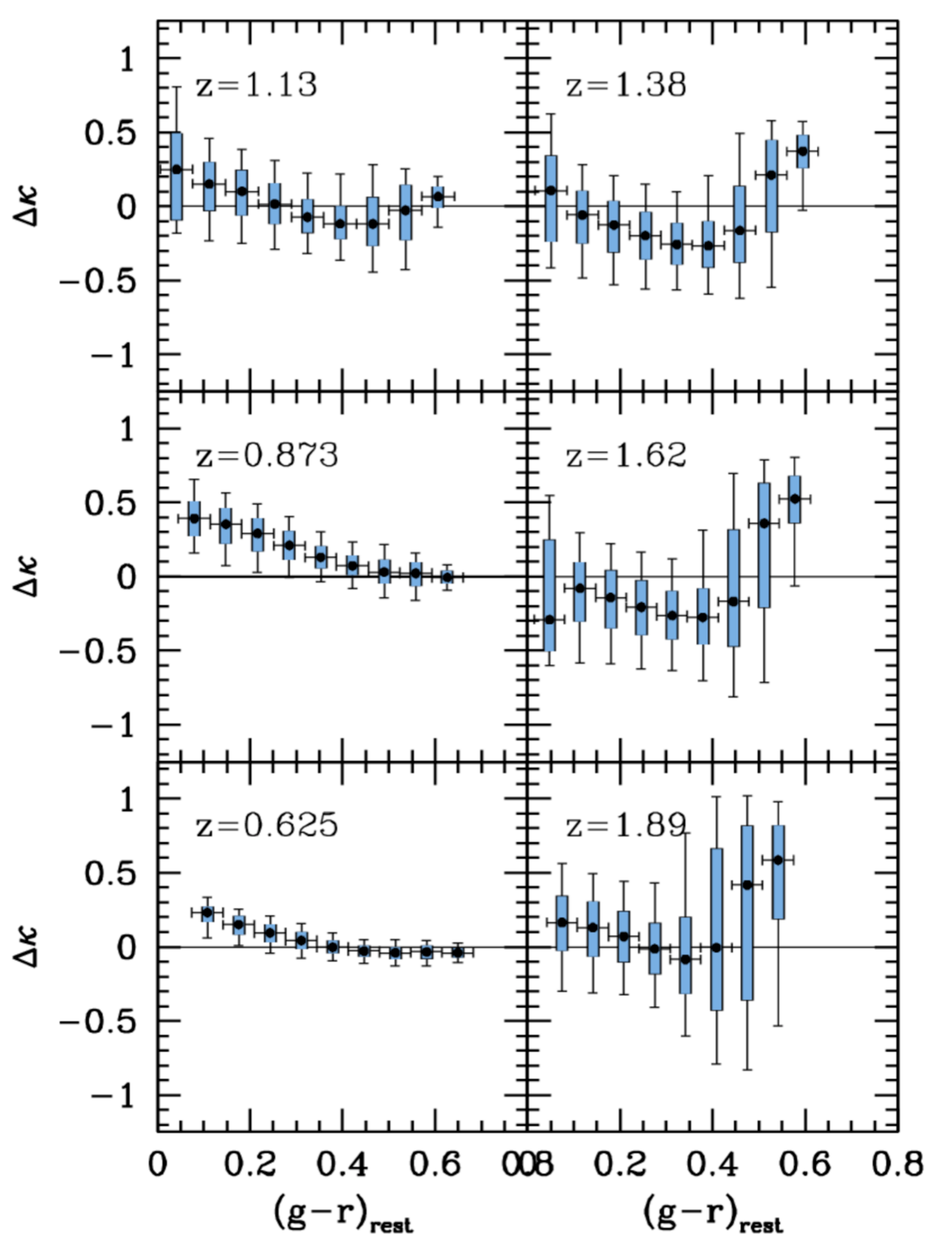}
\caption{Fitting uncertainties of the $r$--band $k$-correction. Symbols locate the median values and boxes show the 1--$\sigma$ dispersion while the error bars indicate the 2--$\sigma$ scatters.}
\label{fig:fit}
\end{figure}

\begin{figure}
\centering 
\includegraphics[width=\linewidth]{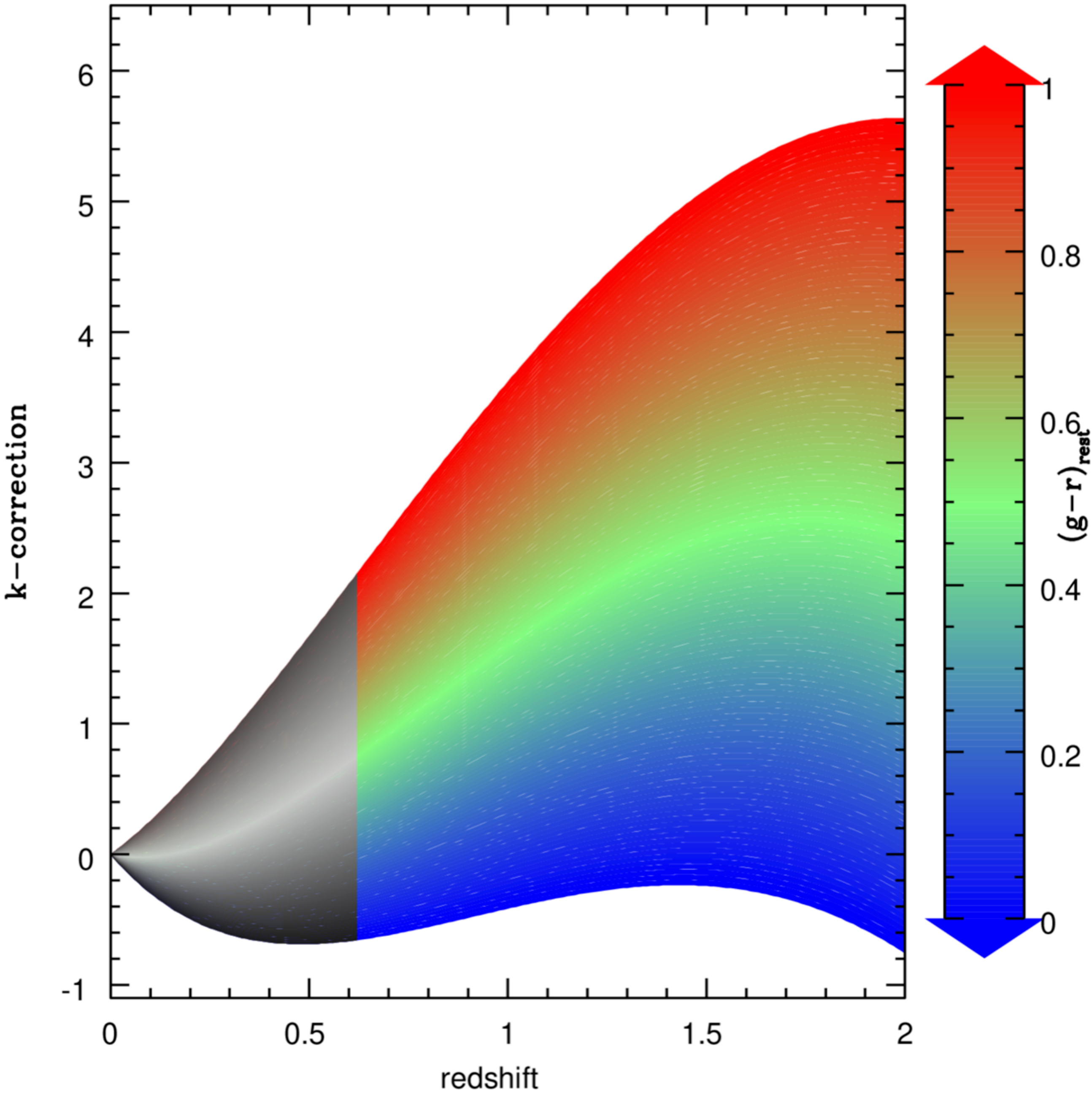}
\caption{The fitting model of the $k$-correction formulated in Eq. (\ref{eq:K-fit}). We colorized the distribution of the modeled $k$-correction for each $(g-r)_{\rm rest}$ as shown in the right scale bar. Note that the valid redshift range for the fit is $0.625\le z \le 2$, but we extrapolate the fit down to $z=0$, in the redshift range marked by gray.}
\label{fig:modeled}
\end{figure}

For the simulated galaxies, we may directly shift their rest frame SEDs to a target redshift and measure the absolute magnitudes in the frame as \citep{blanton07}
\begin{equation}
    \mathcal{K} \equiv -2.5 \log\left[ {{1\over (1+z)} { {\int d\lambda_0 \lambda_0 L_\lambda (\lambda_0/1+z) R(\lambda_0)} \over {\int d\lambda_0 \lambda_0 L_\lambda (\lambda_0) R(\lambda_0)}}} \right]
\end{equation}
where $R$ is the filter response function and $L_\lambda$ is the rest-frame SED of the galaxy. We have found that the $k$-correction can be formulated as a function of the rest-frame color and redshift.
The best-fit model for the $k$-correction in $r$-band is
\begin{equation}
\hskip -0.3cm	\mathcal{K}^{\rm fit}_{r} \!=\! \left\{13.3(g\!-\!r)_{\rm rest}\!-\!0.5\!-\!3.5(z\!-\!1.47)^2\right\} \log(1\!+\!z),
	\label{eq:K-fit}
\end{equation}
for the redshift range of $0.625\le z \le 2$. Figure \ref{fig:fit} shows the fitting results at several redshifts. The $k$-correction difference is defined as $\Delta \mathcal{K}\equiv \mathcal{K}_r^{\rm sim}-\mathcal{K}_r^{\rm fit}$, where $\mathcal{K}^{\rm sim}$ is directly derived from the simulated SEDs. They seem to have a small non-linearity in the dependence on the color. Also in Figure \ref{fig:modeled}, we show the redshift evolution of $k$-correction as a function of the galaxy rest-frame color. This figure shows that a smaller $k$-correction is required for bluer galaxies in the $(g-r)_{\rm rest}$ color space.

It is worthwhile to note that Equation~\ref{eq:K-fit} introduces the logarithmic form of redshift rather than polynomials of $z$ \citep[c.f.,][]{OMill+11}, to avoid a linear increase of $\mathcal{K}(z)$ at high redshifts. Moreover, the $k$-correction of Equation~\ref{eq:K-fit} may be a linear function of $\log(1+z)$ with a slope depending on galaxy colors or spectral ages \citep{Stabenau08}. For the typical range of galaxy colors ($-0.2\lesssim {(g-r)}\lesssim 2$), we have $-0.6 \lesssim \mathcal{K}^{\rm fit}/\log(1+z) \lesssim 6$, which is similar to the observational result derived by \cite{Stabenau08}, who measured the change of the surface brightness of galaxies with various types of samples up to $z\sim 5$. 
We use this modeled $k$-corrections to quantify the contribution of the $k$-correction to the GSMFs of galaxies of mock surveys.

\subsection{Cosmological Surface Brightness Dimming Effect}
The stellar mass of observed galaxies can be derived from their luminosities and colors based on the stellar population models and IMFs, while it is intrinsically given from the sum of stellar particle masses in simulations. Thus, it has been suggested that the discrepancy between the GSMFs of simulations and observations in a low-mass end can be eased with the assumption that a substantial fraction of small low-surface-brightness (LSB) galaxies are missed due to the SB limit of observations~\citep{geller+12,Driver_1999,Tang+21}.

In a flat universe, the half-mass angular radius of a galaxy is given by 
\begin{equation}
	\theta^{\arcsec}_{1/2} \simeq 206,265{ r_{1/2} \over d_c (z)},
\end{equation}
where $r_{1/2}$ is the half-mass radius in the proper unit and $d_c$ is the comoving distance to the galaxy, which is a function of the cosmological parameters via
\begin{equation}
	d_c(z) = {c\over H_0}\int_0^z {dz \over E(z)},
\end{equation}
where $E(z) \equiv \{{\Omega_m^0(1+z)^3+\Omega_k (1+z)^2 + \Omega_\Lambda}\}^{1/2}$ in the $\Lambda$CDM model. 
 Now we set the average SB of a galaxy to be the brightness averaged within its half-mass radius ($r_{1/2}$). Note that $r_{1/2}$ is often called as the effective radius.  The surface brightness of a galaxy in the unit of mag arcsec$^{-2}$ is
\begin{equation} 
\hskip -0.2cm	\left<\mu_{1/2}\right> \!=\! \mathcal{M}_\odot \!-\! {5\over 2}\log \left[{L_{1/2}/d_{\rm c}^2(1+z)^2\over L_\odot/(10{\rm pc})^2}\right]\! +\!{5\over 2}\log\left(\pi\theta_{1/2}^{\arcsec2}\right), \label{eq:sb_org}
\end{equation}
where $\mathcal{M}_\odot$ is the absolute magnitude of the Sun, $L_\odot$ is the solar luminosity, and $(1+z)^2$ is inserted to reflect the cosmological expansion. The first two terms in the right-hand side of Equation~\ref{eq:sb_org} is the observed magnitude of the galaxy, while the last term is added to convert the observed magnitude into average surface brightness per square arcsecond. 

For an observation adopting a filter ($X$) of a finite bandwidth, Equation~\ref{eq:sb_org} needs to be modified into
\begin{eqnarray} \nonumber
	\left<\mu_{1/2}^X\right> &\simeq&\mathcal{M}_\odot^X - {5\over 2}\log \left({\mathcal{L}_{1/2}\over \mathcal{R}_{1/2}^2}\right) + {5\over2}\log(1+z)^3 \\
	&& +\mathcal{K}(z;X) + 37.82, \label{eq:sbm}
\end{eqnarray}
where $\mathcal{K}$ is the $k$-correction term for the redshifted spectral energy distribution (SED) of the galaxy, $\mathcal{L}_{1/2}\equiv L_{1/2}^X/L_\odot^X$, and $\mathcal{R}_{1/2}\equiv r_{1/2}/1{\rm kpc}$, where $r_{1/2}$ is the half-light radius in a proper scale.
Equation~\ref{eq:sbm} can be simplified as 
\begin{equation} \label{eq:mu_org}
	\left<\mu_{1/2}^X\right>	 \simeq \mathcal{M}^X + 5\log\mathcal{R}_{1/2}+\mathcal{C}(z;X) + 38.57,
\end{equation}
where $\mathcal{M}^X$ is the absolute magnitude of a galaxy and $\mathcal{C}(z;X)$ is the cosmological expansion term:
\begin{equation} \label{eq:mu_cosmos}
	\mathcal{C}(z;X) \equiv \mathcal{K}(z,X) + {5\over2}\log (1+z)^3,
\end{equation} 
which depends only on redshift, irrespective of the cosmological model parameters~\citep{Stabenau08}. When the cosmological factor is not included, we call it the rest-frame surface brightness as
\begin{equation}
	\left<\mu_{1/2}^{\prime X}\right>	 \simeq \mathcal{M}^X + 5\log\mathcal{R}_{1/2}+ 38.57.
\end{equation}

If we neglect the expansion effect, the surface brightness of a galaxy is invariant irrespective of the distance from an observer.

\subsection{Distribution of the Surface Brightness of Galaxies}
\begin{figure}
\centering 
\includegraphics[width=\linewidth]{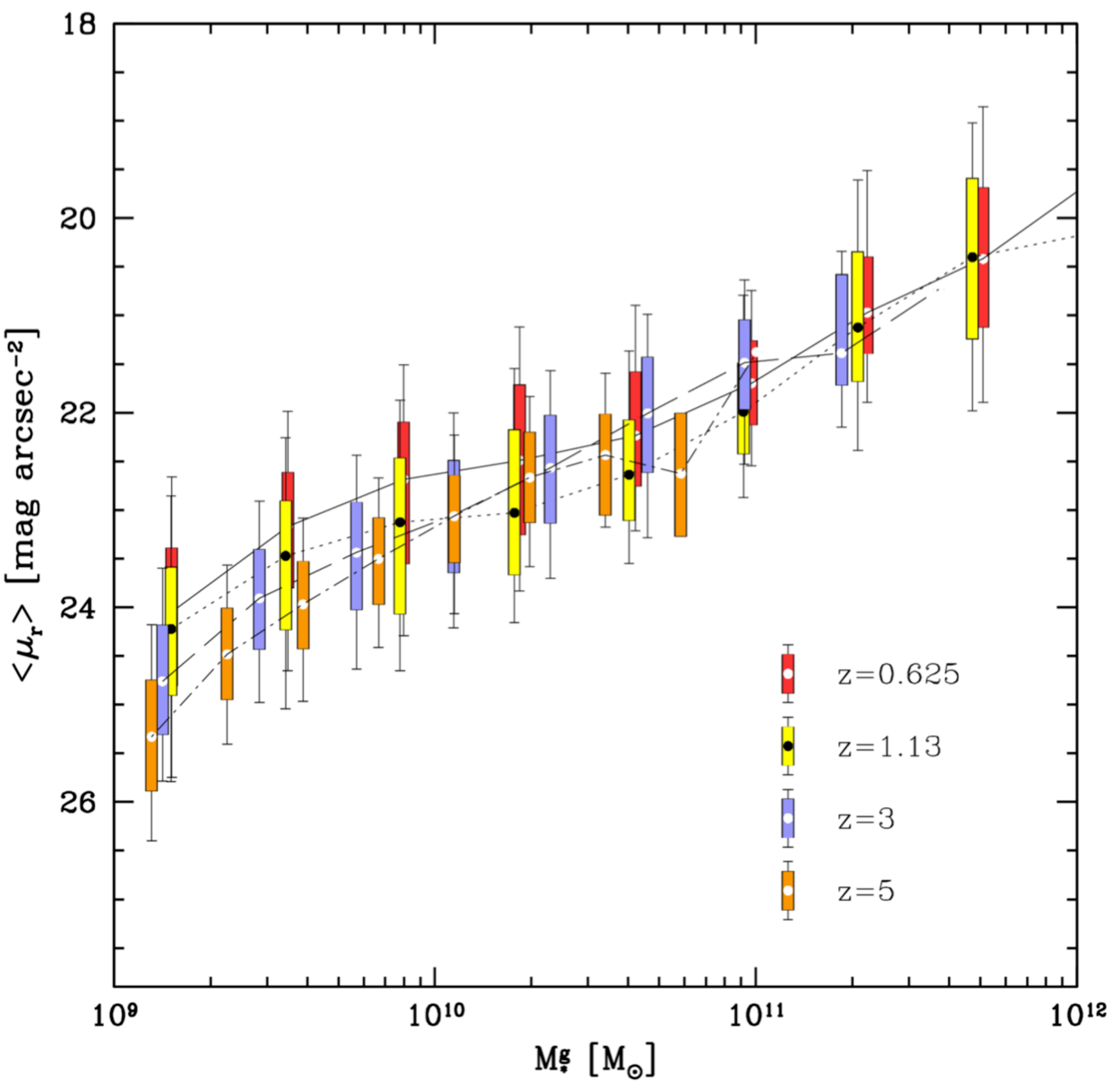}
\caption{Distributions of mean SB of the \hr\ galaxies that are assumed to be observed at $z=0$. The filled circle with error bars marks the median value and the 2--$\sigma$ (95\%) scatter while the colored box ranges the 1--$\sigma$ (68\%) distribution of the mean surface brightness.}
\label{fig:muhist}
\end{figure}

We calculate the mean SB of the simulated galaxies of \hr\ using Equation~\ref{eq:sbm} at four redshifts of $z=0.625$, 1.13, 3, and 5. In each stellar mass bin, we measure the median, 1--$\sigma$, and 2--$\sigma$ distributions as shown in Figure \ref{fig:muhist}. More massive galaxies tend to have higher SB because the increase of brightness or stellar mass outpaces the increase of the projected surface area. Although galaxies are more compact and brighter at a fixed mass at higher $z$, the SB is higher at lower $z$ due to thecosmological expansion term $\mathcal{C}(z;X)$ in Equation~\ref{eq:mu_cosmos}. Accordingly, for instance, a galaxy redshift survey with the target SB limit of $\left<\mu_r\right>$=26 mag arcsec${}^{-2}$ will be able to build a complete catalog of galaxies more massive than about $10^{11}{\rm M_\odot}$ in the redshift range of $0.7\lesssim z \lesssim 2$. However, if the SB limit is lower than e.g., $\left<\mu_r\right>$=24 mag arcsec${}^{-2}$, the galaxy survey would only cover $M_\star\gtrsim 10^{12} {\rm M_\odot}$ in the redshift range. 

\begin{figure}
\centering 
\includegraphics[width=\linewidth]{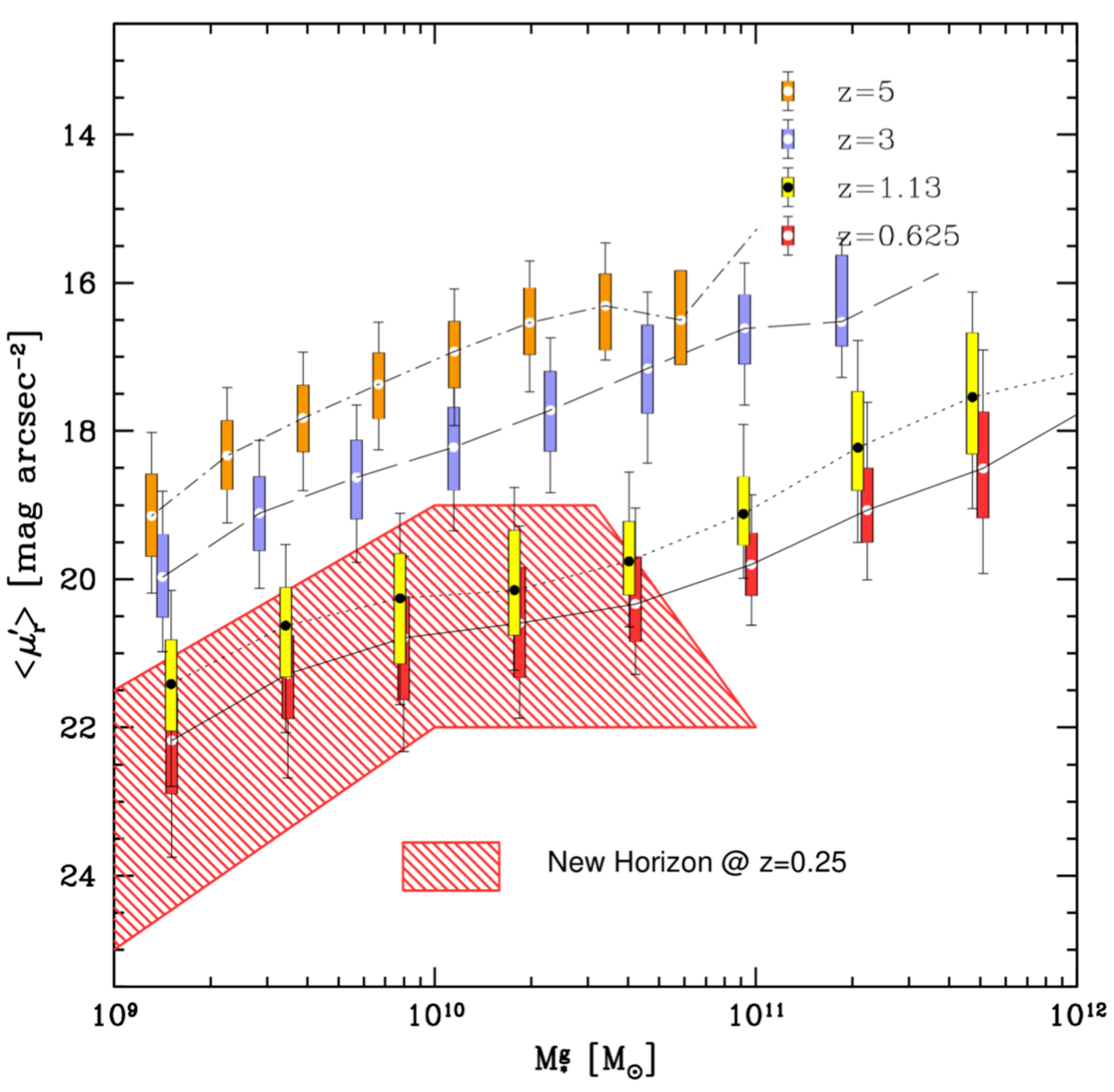}
\caption{The rest-frame of the \hr\ galaxies obtained by using Equation~\ref{eq:mu_org} without the terms depending on the cosmic expansion. The mean SB versus stellar mass relation of the New Horizon simulation ~\citep{jackson20} at $z=0.25$ is included for comparison.}
\label{fig:mu_noc_hist}
\end{figure}

Figure~\ref{fig:mu_noc_hist} demonstrates the impact of the cosmological expansion term $\mathcal{C}(z;X)$ in Equation~\ref{eq:mu_org} on the SB of galaxies. In this case, the SB is only governed by the evolution of astrophysical parameters: effective radius and the absolute magnitude. Because of their compact size and younger stellar populations, higher-$z$ galaxies have notably higher SB.

\subsection{Completeness of GSMF by The SB Limit}

\begin{figure}
\centering 
\includegraphics[width=\linewidth]{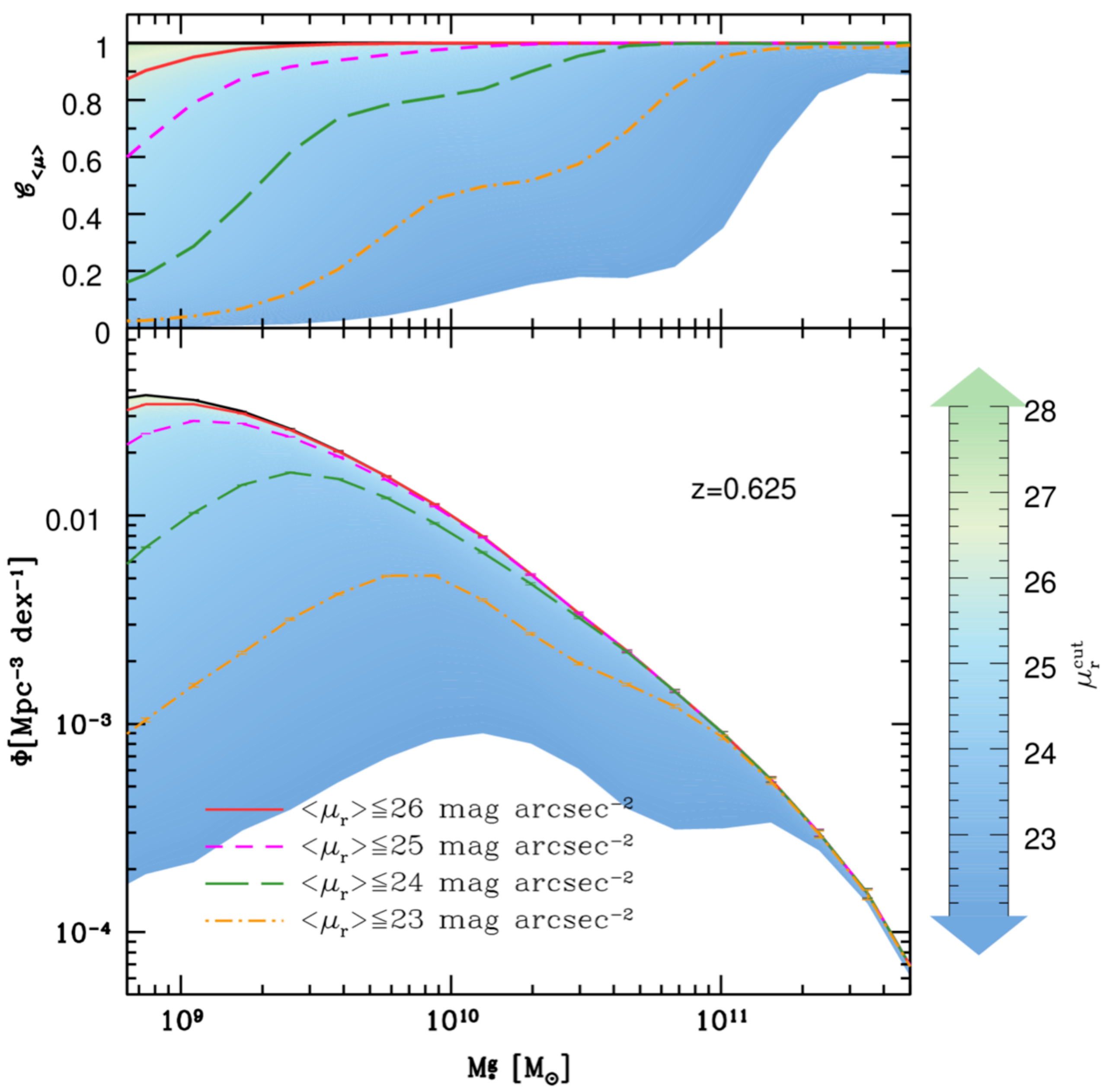}
\caption{Mock observations of the GSMFs with various surface-brightness limits at $z=0.625$. ({\it bottom}) The black solid line is the stellar mass functions with all galaxies while the other lines are obtained with surface brightness cuts. Error bars mark the 1--$\sigma$ Poisson errors. ({\it top}) The dependence of the completeness of mock samples on the input surface-brightness cut. Note that in this plot we do not apply the 30 ckpc aperture correction to the stellar mass.}
\label{fig:statSB}
\end{figure}

Now, we examine the effect of the SB detection limit on mock galaxy surveys. Mock galaxy surveys are made with various SB limits without implementing any apparent magnitude limit, to isolate the SB limit effect. Therefore, this is an extreme case of the extragalactic surveys, where the SB of a galaxy is the only constraint on source detection. Figure \ref{fig:statSB} shows the effect of SB limit on the GSMF at $z=0.625$. As expected, the SB limit has more impact in low mass galaxies; dwarf galaxies are missed more with a lower SB limit. The vertical axis ($\mathscr{C}_{\rm sb}$) of the top panel is the survey completeness defined as
\begin{equation}
	\mathscr{C}_{\left<\mu_r\right>} \equiv {N_{\rm obs}\over N_{\rm all}},
\end{equation}
where $N_{\rm obs}$ is the number of galaxies which satisfy the SB condition of the mock survey and $N_{\rm all}$ is the total number of target galaxies.
\begin{figure}
\centering 
\includegraphics[width=\linewidth]{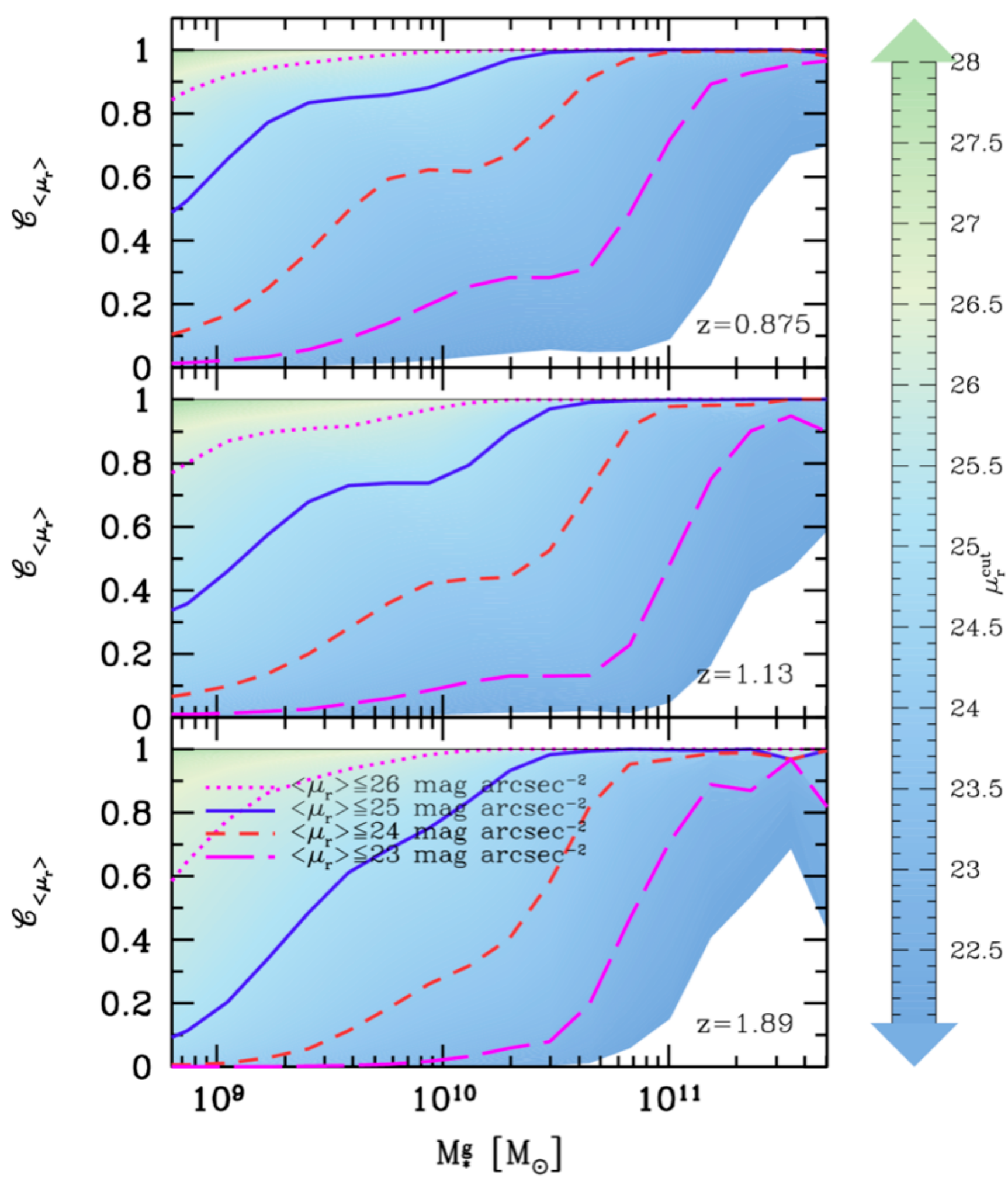}
\caption{Completeness of mock observations made with several surface brightness cuts (values in the legend). From the top panel, the survey completeness (curves) is shown in a mock survey of galaxies at $z=0.875$, 1.13, and 1.89.}
\label{fig:SBhist}
\end{figure} 

In Figure \ref{fig:SBhist}, we show the completeness of the GSMF in mock observations at $z=0.875$, 1.13 and 1.89 (from the top panel) as a function of the SB limit. For example, we may fail to detect 35 \% of galaxies with $M^g_\star=10^9~{\rm M_\odot}$ if the SB limit of the survey is $\left<\mu_r\right>= 25$ mag arcsec$^{-2}$ at $z=0.875$. The fraction of missing LSBGs inevitably increases with increasing redshifts.

\begin{figure}
\centering 
\includegraphics[width=\linewidth]{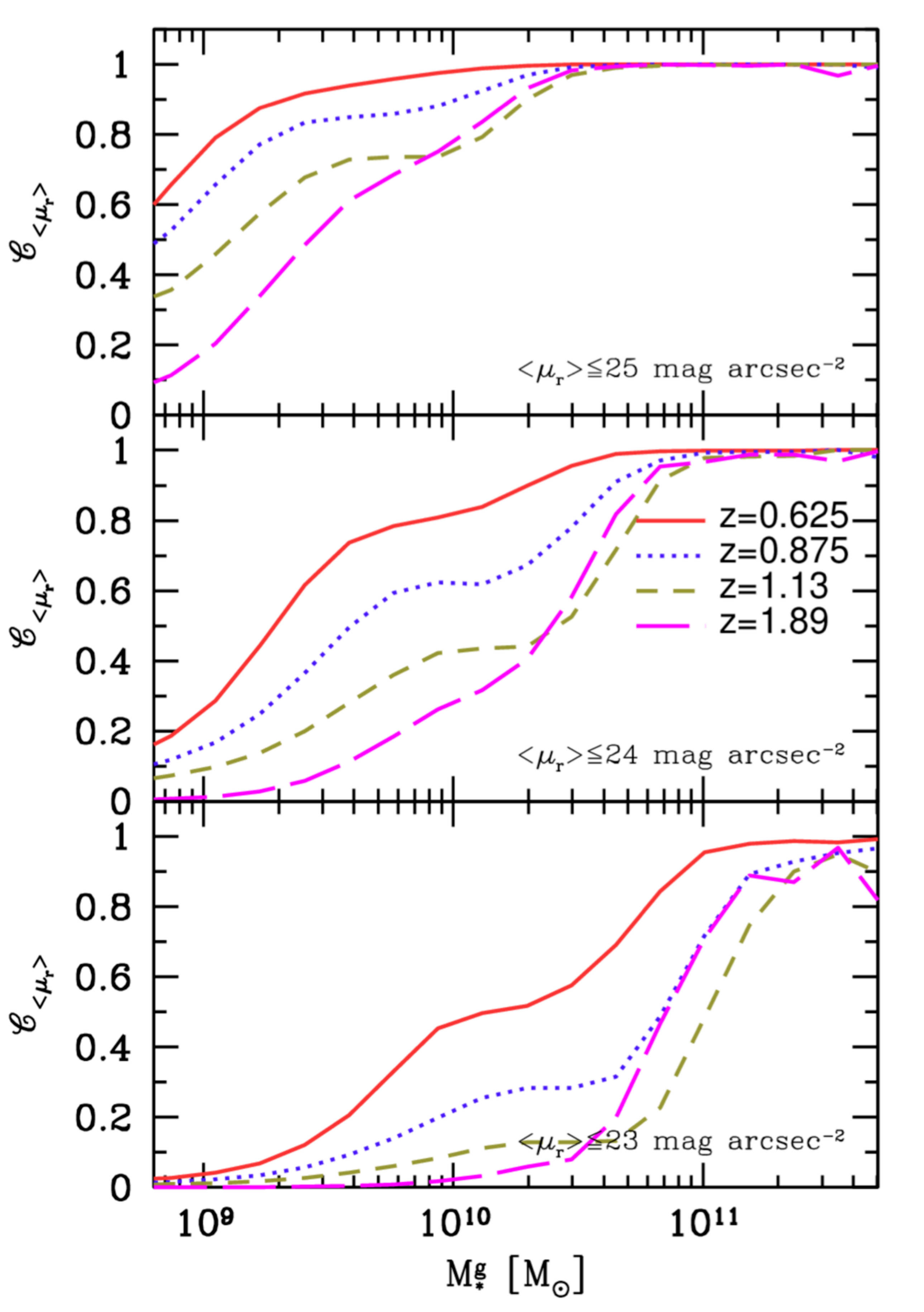}
\caption{Redshift evolution of the fraction of observationally selected galaxies for the mean surface brightness cut of $\left<\mu_r\right>$=25 (top), 24 (middle), and 23 mag arcsec${}^{-2}$ (bottom panel).}
\label{fig:SBhist.evol}
\end{figure}

It is valuable to note that, in spite of strong star formation activities at high redshifts, the cosmological SB dimming effect becomes dominant, eventually lowering the SB of galaxies, as shown in Figure \ref{fig:SBhist.evol}. This figure suggests that the GSMF at $z\lesssim 2$ would be complete down to $M=10^{11}{\rm M_\odot}$, if a survey adopts the SB condition of $\left<\mu_r\right> \leq24$ mag arcsec${}^{-2}$. 

%A caveat of this analysis is that \hr\ has a spatial resolution of $\sim1\,$pkpc which is comparable or larger than the half-mass radius of galaxies in the mass range of $M_\star^g\sim10^9-10^{10}\msun$ (see the red solid line and the thin blue shade in Figure~\ref{fig:galsize}). Thus, the size of low mass galaxies can be overestimated in \hr. Figure~\ref{fig:galsize} implies that low mass quiescent galaxies in \hr\ may have sizes twice larger than observations, on average, which leads to 1.5 mag arcsec$^{-1}$ fainter SB. Therefore, depending on the morphology distribution of the simulated galaxies in \hr, the missing fraction of LSBGs given by the SB limit of $\left<\mu_r\right>= 26.5$ mag arcsec$^{-2}$ may be close to that given by the SB limit of $\left<\mu_r\right>= 28$ mag arcsec$^{-2}$.

%

\begin{figure*}
\centering 
\includegraphics[width=\linewidth]{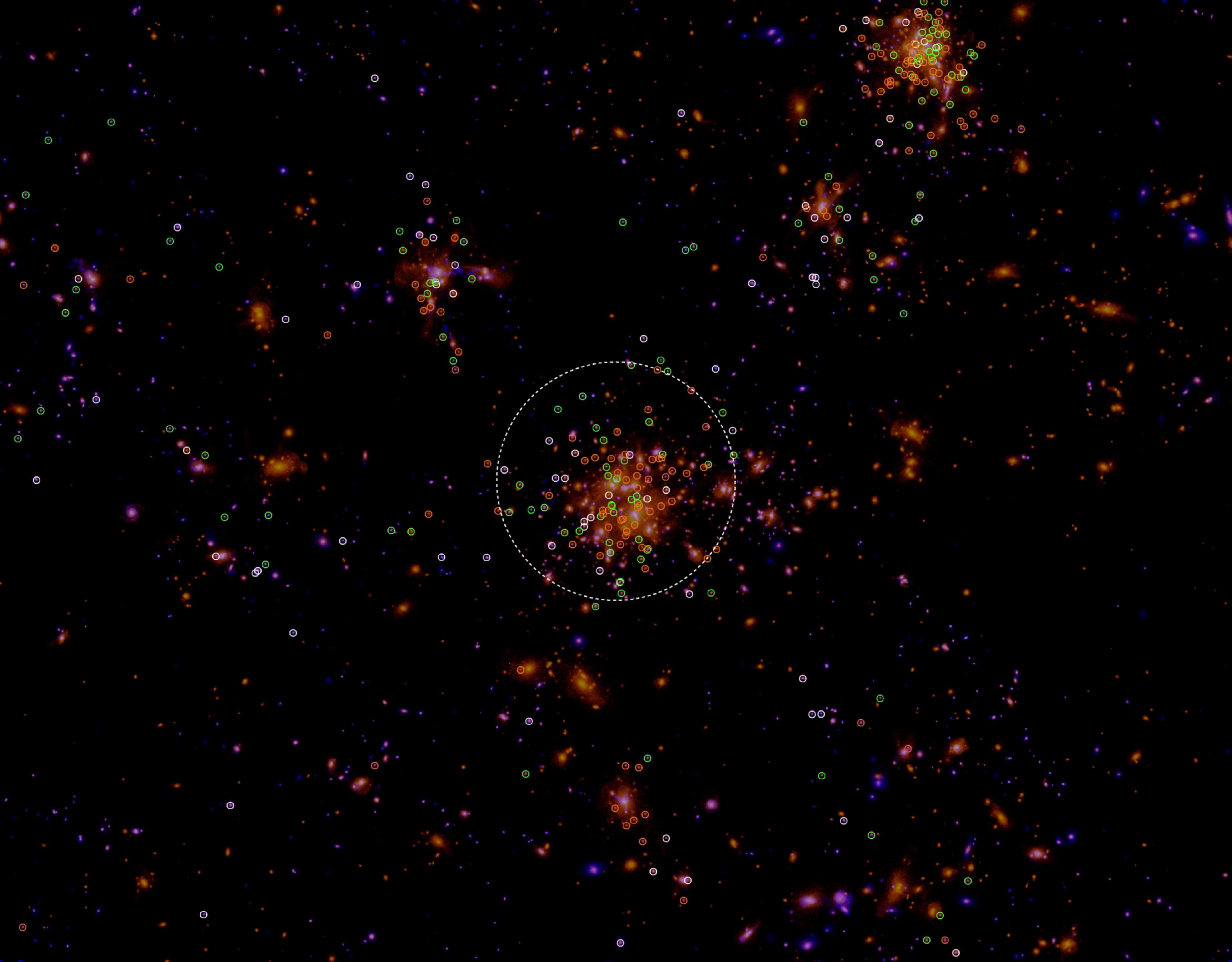}
\caption{A pseudo color map of the star+gas density field of \texttt{Cluster} 1 of \hr\ at $z=0.625$. Open circles are the LSBGs located within 10~cMpc from the cluster center that have stellar mass of $10^9 {\rm M_\odot} \le M_\star\le 10^{10}{\rm M_\odot} $. The white, green, and red circles with $r_\theta=4{\arcsec}$ mark the LSBGs in the SB ranges of [23, 24), [24, 25), and $[25,\infty)$  mag arcsec${}^{-2}$, respectively. Note that the most of unmarked small faint blobs scattered around the cluster are LSBGs located beyond   10 $h^{-1}$cMpc from the cluster center. The dotted  circle has the radius of $R=1~h^{-1}$ cMpc.}
\label{fig:lsbimg}
\end{figure*}

Figure~\ref{fig:lsbimg} presents an example of the distribution of LSBGs identified around \texttt{Cluster} 1 region in \hr\ at $z=0.625$~\citep{lee21}. The LSBGs located within a radius of 10 $h^{-1}$ cMpc from the cluster's center are marked by the open circles with colors for various SB ranges. The LSBGs are faint but distinct with compact sizes, without prominent tidal features. There are two groups of galaxies falling into the cluster in the upper right and left parts (around 1 \& 11 o'clock) of the image. The group also contains several LSBG satellites. In this LSBG distribution, we can see that the cluster LSBGs scatter around the central part of the cluster, and some of them stretch along the local filament connecting the cluster to neighboring galaxy groups.

\section{Galaxy Stellar Mass Functions at $z\gtrsim 2$}
\label{sec:4}
The GSMF is one of the fundamental statistics that galaxy formation models aim to reproduce. Since the cosmic dawn, the average star formation density of the universe has increased through $z=2$--3. Then, star formation is suppressed particularly in massive galaxies as AGNs start to emit sufficient energy that blows away or heats up the interstellar medium. The global SFR declines until today, with many implications imprinted on the GSMF, galaxy morphology, and color distributions. The shape of the GSMF reflects a sophisticated history of cosmic matter evolution. For example, the slope above the knee of a Schechter function is known to be suppressed by energetic AGN feedback in massive galaxies that is controlled by the accretion of gas via mergers or secular evolution. %important factors involve the gravitational coupling of dark matter and baryonic matter in an expanding medium, gaseous dissipation, SN and AGN feedback, and galaxy mergers, each of which intertwines with the others, driving or suppressing star formation activities.

\subsection{\hr\ versus \texttt{JWST} at High Redshifts}
Our universe still has much to discover and remains open to new discoveries. Recently, \texttt{JWST} has observed galaxies in their infant era beyond $z=10$, which will give us a hint on the early structure formation and enable us to test the standard model of cosmology and galaxy formation. 

\begin{figure}
\centering 
\includegraphics[width=\linewidth]{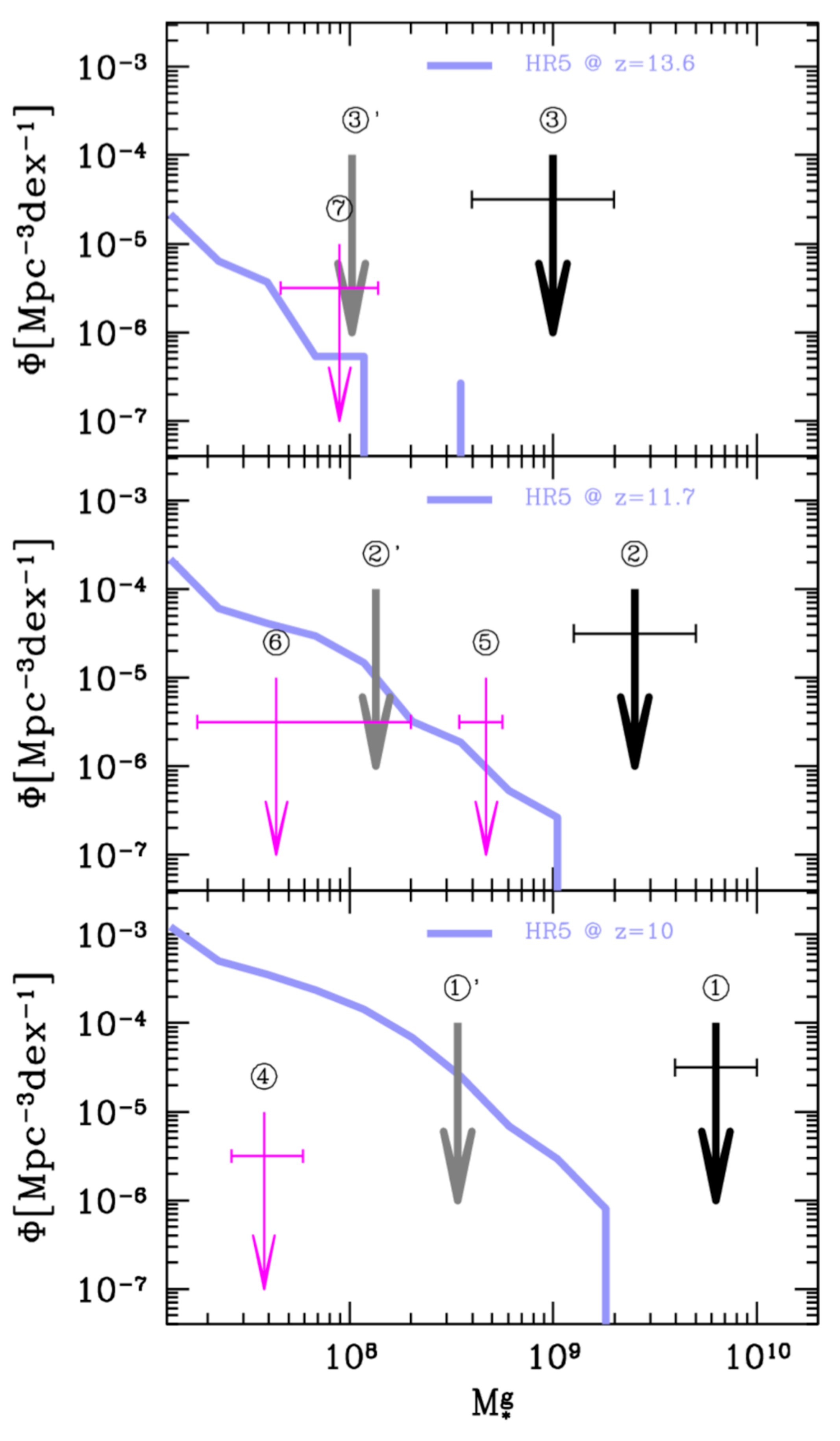}
\caption{Comparison with \texttt{JWST} observations. From the top panel, we show the \hr\ mass functions (thick line) at $z=13.6$, 11.7, and 10, respectively.  ({\it observations}) \cite{haslbauer+22}: \numcircledmod{1} is ID 1514 ($z=9.9$), \numcircledmod{2} is GL-z11 ($z=11$), \& \numcircledmod{3} is GL-z13 ($z=13$). \cite{curtis-lake+22}: \numcircledmod{4} is GS-z10-0 ($z=10.38$), \numcircledmod{5} is GS-z11-0 ($z=11.5$), \numcircledmod{6} is GS-z12-0 ($z=12.63$), \& \numcircledmod{7} is GS-z13-0 ($z=13.2$). The gray downside arrow marks the galaxy stellar mass calculated when an environment-dependent IMF is applied \citep[for details, see][]{haslbauer+22}. }
\label{fig:jwst}
\end{figure}

\cite{haslbauer+22} argued that the \texttt{JWST} observation may falsify the standard $\Lambda$CDM model due to the relatively massive galaxies found at high redshifts compared to the cosmological hydrodynamical simulations such as \texttt{EAGLE} and \texttt{IllustrisTNG}. However, the farthest galaxy candidate used in their analysis is now suspected to be a dusty galaxy at lower-$z$~\citep{naidu+22}. Therefore, we do not use the galaxy sample of \citet{haslbauer+22} in this comparison. In Figure~\ref{fig:jwst}, the black downside arrows mark the upper limit of galaxy number density at the scale of galaxy stellar mass derived from the \texttt{JWST} observations~\citep{haslbauer+22}. The mass scale marked by the black error bars is lowered to the mass scale marked by the gray error bars if an environment-dependent IMF is adopted. This figure shows that \hr\ is consistent with the \texttt{JWST} observations at $z\simeq10$, depending on the IMF. It is noteworthy that the results are based on the photometric redshift, and thus they should be treated with caution. However, the redshifts of the galaxy sample in the JWST Advanced Deep Extragalactic Survey (JADES) are obtained from the spectroscopic observations~\citep[in magenta downside arrows][]{curtis-lake+22}. If the simulation volume is bigger by an order of magnitude, galaxies could properly be simulated on the observed mass scale~\citep[also see][for other possible factors]{keller+23}. On the other hand, \citet{donnan+23} confirm that the galaxy UV luminosity function from \texttt{JWST} does not exceed the $\Lambda$CDM prediction significantly. We may conclude with \hr\ that the high-$z$ observations are consistent with the standard hierarchical structure formation model, so far, as illustrated in the figure.

\subsection{GSMF at $5 \le z \le 10$}
Galaxy observational data compiled over the last two decades in many legacy surveys have enabled us not only to trace the global star formation and reionization history in the early universe, but also to study proto-galaxy formation with a high precision. In this study we utilize various observational data~\citep{ Gonzalez_2011,Duncan_2014,Grazian_2015,Song_2016,stefanon21} which have been obtained from the GOODS and CANDELS fields observed by HST and {\it Spitzer} IRAC to name a few.

\begin{figure*}
\centering 
%\plotone{redhigh}
\includegraphics[width=1\linewidth]{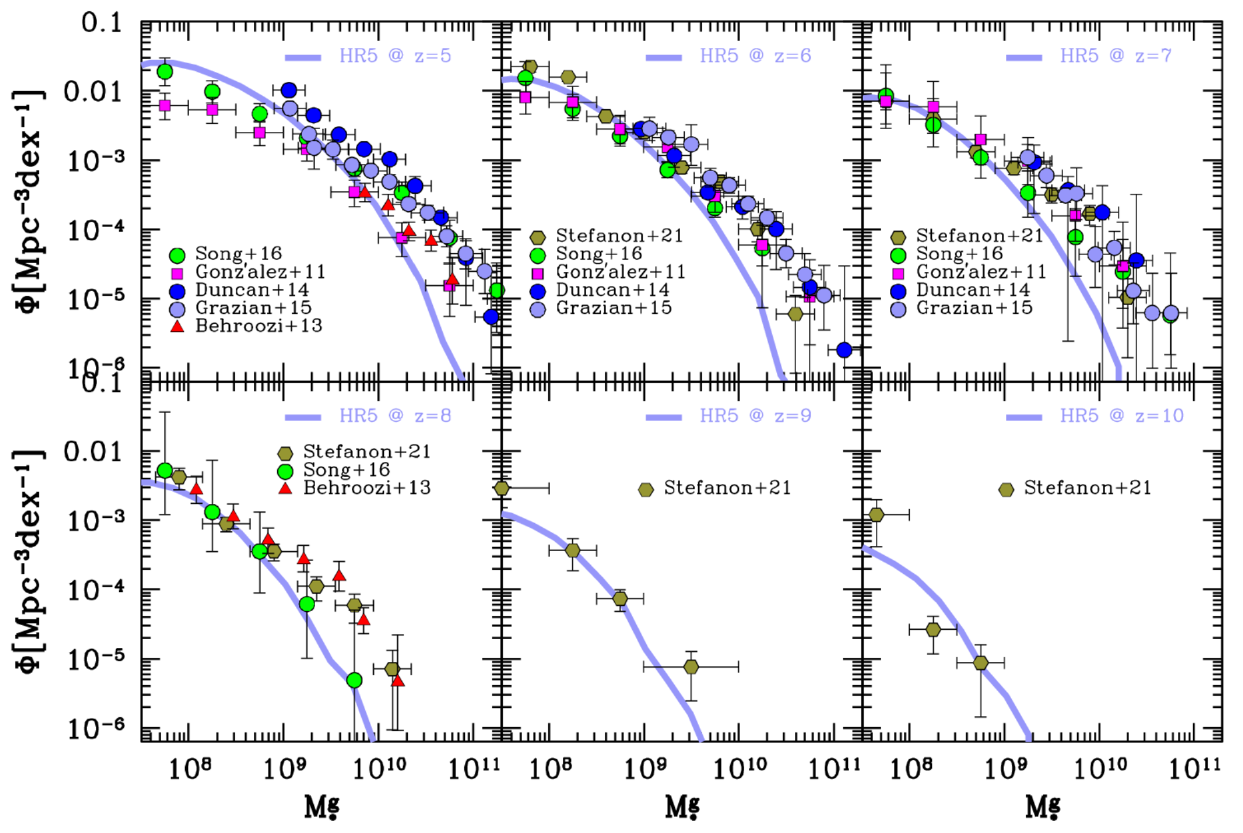}
\caption{Comparison of GSMFs at high redshifts. Symbols with error bars  are the observations (see the main text for references) at $z=$ 5, 6, 7, 8, 9, and 10 (from the top-left to the bottom-right panel in the sinistrodextral order) while solid lines are the corresponding simulated stellar mass functions of \hr.
{\it Observations}: \citet{Song_2016}, \citet{Gonzalez_2011}, \citet{Duncan_2014}, \citet{Grazian_2015}, and \citet{behroozi13}, and \citet{stefanon+21}. }
\label{fig:redhigh}
\end{figure*}

In Figure \ref{fig:redhigh}, the simulated GSMFs (blueish solid lines) are consistent with observations, except for the high-mass end at $z<7$. This deviation partly comes from the cosmic variance effect caused by the finite simulation box size of \hr\ or the small star formation efficiency (SFE; see \ref{sec:sfecomp}) adopted in \hr. Even with the small volume size of the test simulations, the GSMFs at $z\ge 5$ are underestimated by 1.5 -- 3 factors with respect to the simulation with SFE=4 \%, which may partially explain the difference as shown in the figure.

It may, however, also be attributed to numerical artifacts of AMR, which causes suppression of star formation between global mesh refinement redshifts. AMR is a classical implementation to enhance the spatial and mass resolution in Eulerian hydrodynamical simulations, in the regions where the mass inflow becomes significant. In \hr, this global refinement is carried at the expansion factors of $a_{\texttt{amr}}$ = 0.0125, 0.25, 0.05, 0.1, 0.2, 0.4, and 0.8, in order to maintain spatial resolution of $\sim1$ pkpc. As the simulation evolves, a dense gas grid inside a massive galaxy must wait for the next $a_\mathrm{amr}$ to be refined into eight one-level-finer grids; hence star formation is suppressed until the next-level refinement is allowed. 

At $z=5$ , the simulated GSMF at low-mass end is slightly higher than the observations \citep{Song_2016,Gonzalez_2011} by a few ten percent. This mismatch could possibly be due to the limit to the observational completeness of galaxy sample. The faint-end galaxies usually lie at the observational boundaries and we assume that faint galaxy samples could suffer from the completeness \citep[see][for the simulated completeness of observations]{finkelstein+15}. The same argument may go to the observed GSMFs at $z=2.5$ and 3.5 in the next subsection.

\subsection{GSMF at $2\lesssim z\lesssim 4$}
\begin{figure*}
\centering 
\includegraphics[width=\linewidth]{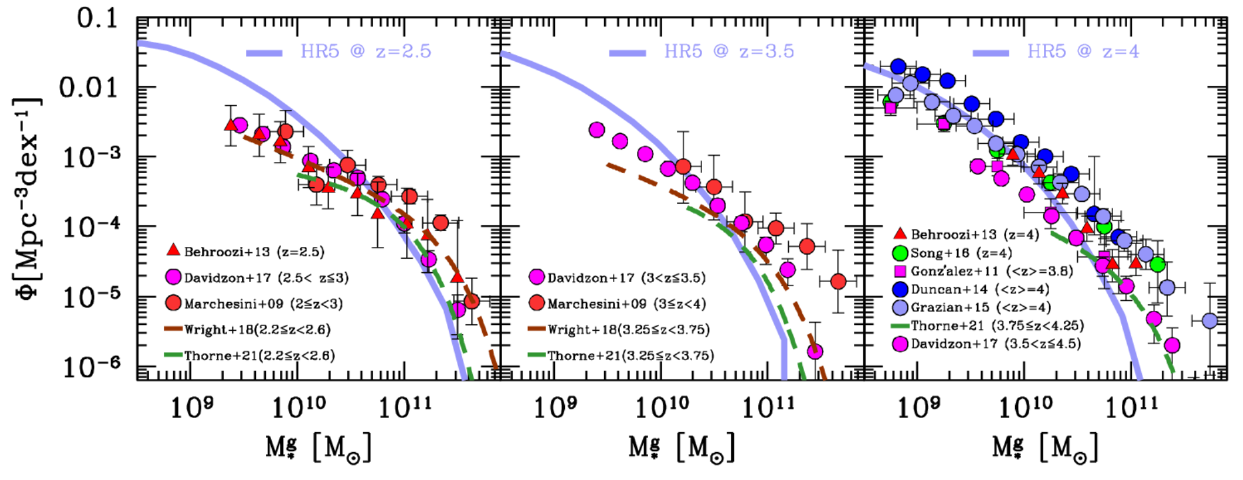}
\caption{Comparison of GSMFs at intermediate redshifts ($z=$ 2.5, 2.5, and 4 from the left panel). The GSMFs of \hr\ are denoted by the thick solid lines while the observed mass functions are marked with dashed lines or symbols with error bars. {\it Observations}: \citet{behroozi13}, \citet{Davidzon_2017}, \citet{Marchesini_2009}, \citet{Wright_2018}, \citet{Thorne+21}, \citet{Gonzalez_2011}, and \citet{Duncan_2014,Grazian_2015}.}
\label{fig:redmid}
\end{figure*} 
In order to carry out a robust comparison of GSMFs between the observational results and simulations at intermediate redshifts, we take five different observed GSMFs~\citep{Marchesini_2009, Davidzon_2017, Wright_2018, Thorne+21, Song_2016} among others (for a full list of references, see the legend in Figure~\ref{fig:redmid}). Figure~\ref{fig:redmid} shows the GSMFs of \hr\ and observations at the three redshifts, $z=4$, 3.5, and 2.5. The \hr\ predictions become more consistent with observations at the massive end after $z=4$ at which global refinement takes place. %This may indicate that the effect of numerical artifacts caused by the AMR method becomes less significant . 
From this trend, it is expected that the simulated BCG populations will overshoot the observations at lower redshifts. Meanwhile, we begin to observe the overproduction of low-mass galaxies at $z\le3.5$. This may also indicate less efficient stellar feedback in \hr, like \texttt{H-AGN}~\citep{kaviraj+17}, or that a different approach is needed to explain this low-mass discrepancy. We address this in the next section.

%Until now, we do not apply the SB dimming effect on the observed galaxy populations because they are obtained from the narrow but deep pencil-beam observations. The SB limit is so high enough to cover the galaxy stellar mass range in this study.

%
\section{Galaxy Stellar Mass Function at $0.625\le z\le 2$}
\label{sec:5}

\subsection{SB Limits of Low-$z$ Observations}
In most galaxy catalogs, the galaxy faint-end magnitude ($m_{\rm ap}^{\rm lim}$) is typically provided, which is determined by the background level and aperture size applied to measure the galaxy total flux. However, full information on the pixel-based source detection is unavailable, which is, on the other hand, crucial to determine the SB limit ($\mu_{\rm p}^{\rm lim}$) of the given galaxy catalog. Due to the limited availability of direct information on SB limits, we must rely on other related information provided in literature. 

Here, we apply some assumptions to derive the SB limit hidden in the galaxy catalog. In Appendix \ref{sec:source_detection}, we show the relation between $m_{\rm ap}^{\rm lim}$ and $\mu_{\rm p}^{\rm lim}$ as a function of the source-detection level ($f_{\rm p}$) above the background noise ($\sigma_{\rm p}$) on pixel scale. According to Figure \ref{fig:sb}, the limited SB of galaxies is about 1--2 magnitudes brighter than $m_{\rm ap}^{\rm lim}$ for the same pixel scale ($l_{\rm p}= 0^{\prime\prime}\!\!.15$) as of \texttt{COSMOS} \citep{mccracken+12} between $f_{\rm p}=3$ and 5.

\subsection{Seeing Effects on Observational SB Limits}
There is a serious impact of seeing conditions on the effective SB measurement when a galaxy effective radius is comparable to or smaller than the FWHM of the point-spread function  \citep[PSF, see Appendix \ref{sec:seeing} for the angular-size comparisons \& Figure 6 of][for the seeing effect]{trujillo+01}. The seeing effect depends on the galaxy profile, ellipticity, and specific PSF models. But the overall observed SB becomes fainter than the true SB of a galaxy and this difference becomes severer by $(\Delta\mu)^{\rm seeing} \gtrsim 0.5$ mag arcsec${}^{-2}$ when the FWHM of the PSF is comparable to or larger than the galaxy angular scale of interest.

Although we may, as a result, be able to roughly estimate the SB limit of each survey catalog, we prefer to use the derived value as a reference to fit the observed mass function and provide the best fitting SB limit in the subsequent subsections.
 
\subsection{How to Fit to Observations}
To fit observed GSMFs, we generate the simulation GSMFs for a range of 23 mag arcsec${}^{-2}\le \mu^{\rm lim}\le$ 25 mag arcsec${}^{-2}$ with a step size of $\Delta \mu=0.1$ mag arcsec${}^{-2}$ and apply the $\chi^2$ fitting for data points at $5\times 10^{9}~{\rm M_\odot} \le M^g_\star \le 10^{11}~{\rm M_\odot}$ . 

For the survey catalogs used in this subsection, 
Table \ref{tab:sb} lists the fitting results (fourth column) with the catalog information on the magnitude limit (second column) and the seeing of the image survey in $r$ band (fifth column). For $\mu^{\rm lim}$ we simply subtract a factor of two from $m_{\rm ap}^{\rm lim}$ assuming $f_{\rm p}= 5$ and $f_{\rm ap}=5$. The most probable cause of the difference between $\mu_{\rm fit}^{\rm lim}$ (third column) and $\mu_{\rm p}^{\rm lim}$ (second column) may come from the seeing effect of $(\Delta\mu)^{\rm seeing}$=0.7 -- 1.1 mag arcsec${}^{-2}$.

\begin{deluxetable}{l l l l l }
\tablecaption{The best-fitting SB limit ($\mu_{\rm p}^{\rm lim}$) for each galaxy catalog in $r$ band for $f_{\rm p}= 5$ and $f_{\rm ap}=5$. \label{tab:sb} }
\tablehead{
\colhead{name} & \colhead{$m_{\rm ap}^{\rm lim}$} & \colhead{$\mu_{\rm p}^{\rm lim}$} & \colhead{$\mu^{\rm lim}_{\rm fit}$} &
\colhead{seeing\tablenotemark{a}} 
}
\startdata
\cite{Adams21} & 26.5 & 24.5 & 23.8 &$0^{\prime\prime}\!\!.6$ -- $0^{\prime\prime}\!\!.8$\tablenotemark{b}  \\
\cite{McLeod+21} & 26.4 & 24.4 & 23.8 & $0^{\prime\prime}\!\!.9$ -- $1^{\prime\prime}$\tablenotemark{c}  \\
\cite{Leja_2020} & 27.0\tablenotemark{f} & 24.8 & 24.1 & $\le 1^{\prime\prime}\!\!.3$ \\
\cite{Wright_2018} & $(23.5)_i$\tablenotemark{g} & - & 23.7 & $1^{\prime\prime}\!\!.05$\tablenotemark{d} \\
\cite{Thorne+21} & 26.9 & 24.9 & 23.8 & $0^{\prime\prime}\!\!.74$ -- $0^{\prime\prime}\!\!.76$\tablenotemark{e} \\
\enddata
\tablenotetext{a}{Full width at half maximum (FWHM)}
\tablenotetext{b}{UltraVISTA; \cite{mccracken+12}}
\tablenotetext{c}{UDS, UltraVISTA \& CFHTLS}
\tablenotetext{d}{\cite{andrews+17}}
\tablenotetext{e}{HSC-SSP \citep{aihara+18...70S...4A} 
\tablenotetext{f}{$f_{\rm ap}=3$ \citep{Laigle+16}}
\tablenotetext{g}{Magnitude limit in $i$ band \citep{andrews+17}}
}

\end{deluxetable}

\begin{figure*}[tp]
\centering 
\includegraphics[width=\linewidth]{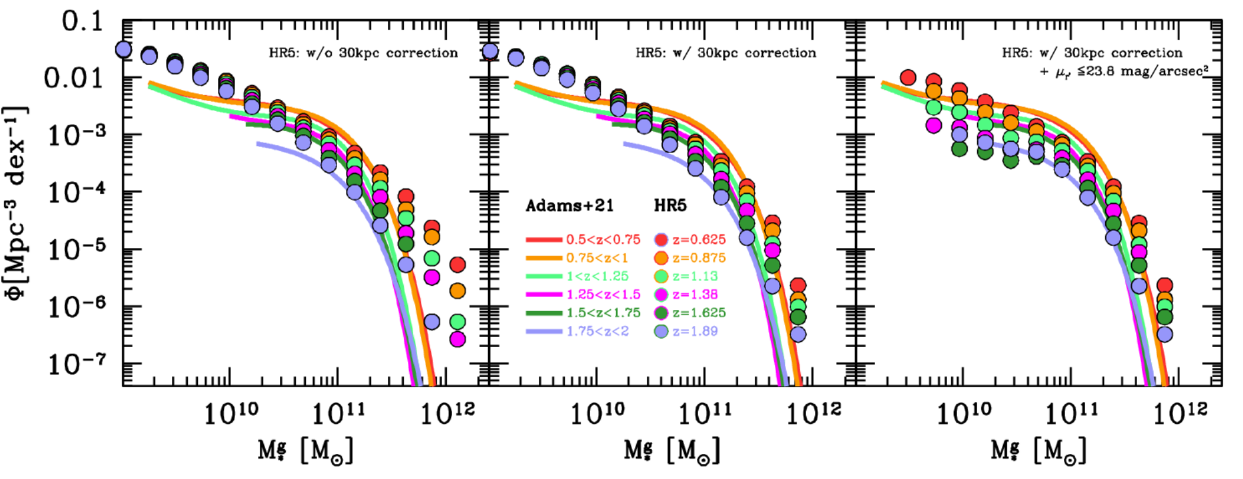}
\caption{GSMFs from the \hr\ ({\it symbols}) and observations (\citealt{Adams21}; {\it lines}) at several lower redshifts. The left panel shows the uncorrected \hr\ GSMF while the middle panel shows the simulation results after applying the 30kpc aperture correction only. The right panel shows the \hr\ GSMF after applying an additional constraint on the surface brightness of galaxies. In this plot we assume that the galaxy data of \cite{Adams21} have a fitting limit SB of about 23.8 mag arcsec${}^{-2}$ (see text for details).}
\label{fig:adams}
\end{figure*} 

\subsection{Results}
Before carrying out a direct comparison with low-$z$ observations, let us briefly summarize the previous findings from observations. There is a consensus that the low-$z$ GSMF evolves slowly under $z\sim 1$ \citep{behroozi13, legrand+19, shuntov+22}. While the GSMF at $1\le z\le2$ grows quickly with a nearly fixed shape, the slope of the GSMF on low-mass scale rises steeply with time \citep{Marchesini_2009, katsianis+15, Leja_2020, Adams21}. 

In contrast, the GSMF of \hr\ is shaped on the dwarf mass scale before $z=2$ and it barely evolves in $z\sim1-2$. On the BCG mass scale, \hr\ shows a strong number-density evolution with decreasing redshift while observations report relatively weak evolution. The \texttt{TNG100} and \texttt{EAGLE} simulations demonstrate the GSMFs that agree well with the observed GSMFs at low redshift since they employ updated versions of the stellar winds and SN feedbacks to fit the simulation GSMF for the low-$z$ observations (see the detailed calibration schemes of \texttt{EAGLE} in \citealt{schaye+15} and \texttt{TNG100} in \citealt{pillepich18}). On the other hand, \hr\ is mainly calibrated to fit the global SFR \citep{lee21}.

In this section, we take two approaches to alleviate these discrepancies based on the fact that the observational techniques for deriving stellar mass from the brightness of galaxies are essentially different to simulations in which galaxy stellar mass is rather intrinsic variables given by stellar particle masses. On the BCG scale, we measure the galaxy stellar mass using finite aperture size to mimic observations. Meanwhile, mainly for the dwarf galaxy scale, the SB of galaxies is taken into account, to understand the shallow slope of GSMFs in a low mass scale in observations~\citep{martin19,williams16}.

%https://arxiv.org/pdf/2009.03176.pdf

Using the galaxy samples in the COSMOS and XMM-LSS fields, together with {\it Spitzer} IRAC, \cite{Adams21} fitted the observed stellar mass functions with the double Schechter functions, in the redshift range of $z=0.2-2$. They corrected the cluster stellar mass with the 30 pkpc aperture size and found that the number density of high-mass end populations drops dramatically compared to the uncorrected ones. 

On the other hand, \cite{Leja_2020} employed a single continuity model for the fitting functions, to take into account the redshift evolution  in the double Schechter function. They find the best fit models by investigating a wider range of parameters. The resulting fits of GSMFs turn out to relax the factor-of-two tension~\citep{Davidzon_2018} between the specific star formation rate (sSFR) and the stellar density by lowering 0.2 dex in the sSFR and raising 0.2 dex in the GSMF. 

The left panel of Figure \ref{fig:adams} shows the empirical fits ({\it lines}; \citealt{Adams21}) and the \hr\ results ({\it symbols}) before applying the 30 pkpc aperture correction. The GSMFs of \hr\ notably deviate from the observations; the over-populations of simulated galaxies are clear on both the high-mass and low-mass ends. However, after applying the 30 pkpc-aperture correction to the simulated galaxies ({\it bottom panel}), we obtain a better correspondence between \hr\ and observations on the BCG scale, recovering the exponential cut-off observed at the high-mass end \citep[also see][]{schaye+15}. However, there are still several differences that should be briefly mentioned here. First, the observation data has a flatter slope below the knees of the GSMFs than those of \hr. Second, the simulation predicts an overpopulation of low-mass galaxies ($M_\star^g \le 10^{10}{\rm M_\odot}$) compared to observations. On the right panel of the figure, we show the effect of a SB limit of $\left<\mu_r\right>\leq 23.8$ mag arcsec${}^{-2}$. Note that the prominent knee of the observed mass function is not reproduced even with the SB limit. This implies that we need to apply more fine tuning to the simulation parameters and the SB limit parameters. However, it is important to acknowledge that the redshift-evolution of dwarf population amplitude is consistent with the observed evolution, which is hardly seen in the original simulation outputs presented in the left panel of the figure.
\begin{figure}
\centering 
\includegraphics[width=\linewidth]{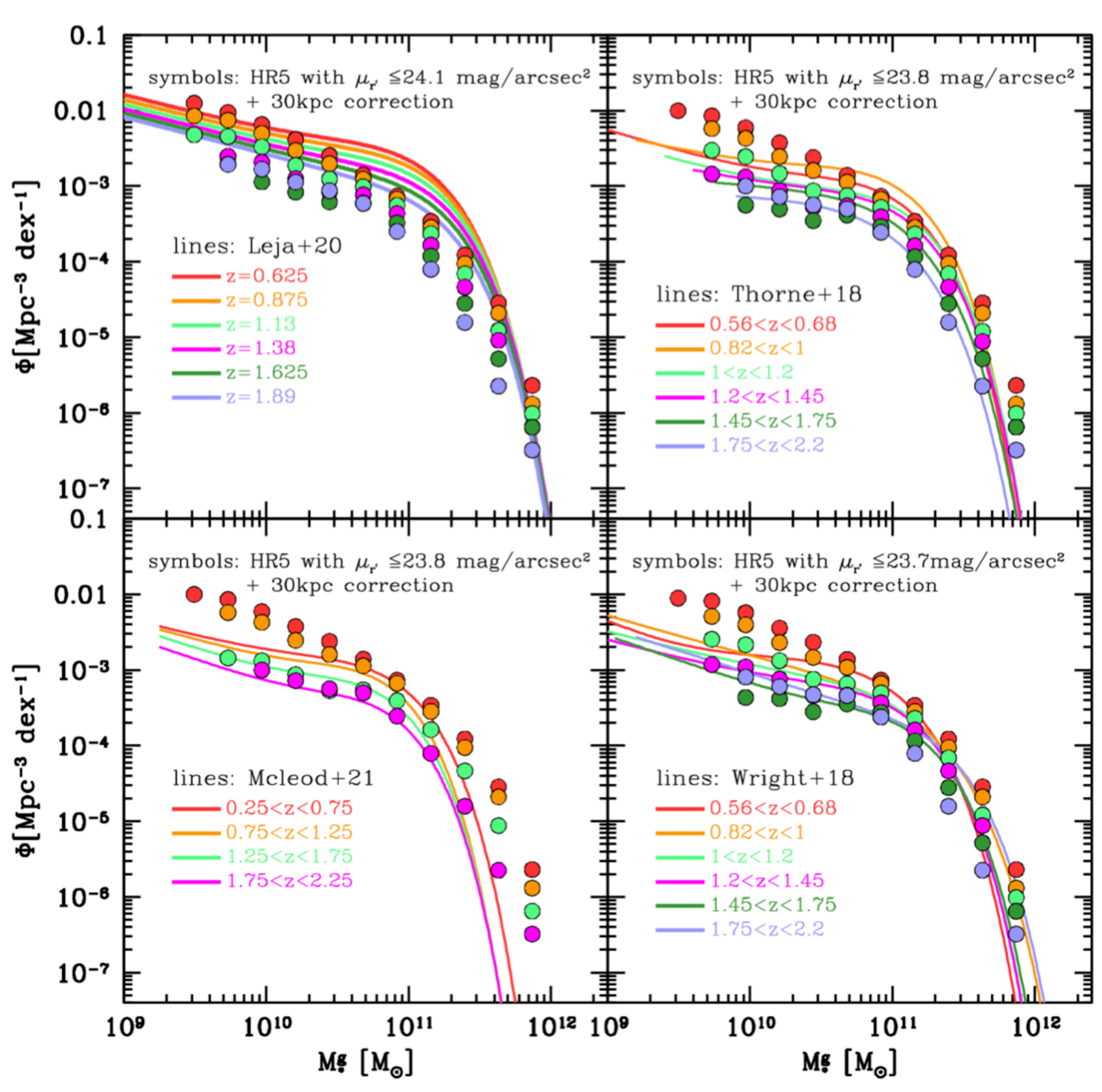}
\caption{Comparisons of GSMFs between the \hr\ ({\it symbols}) and observations. Counterclockwise from the upper-left panel, the lines are the observed GSMFs given by \cite{Leja_2020}, \cite{McLeod+21}, \cite{Wright_2018}, and \cite{Thorne+21} at several lower redshifts. Each color indicates the same redshift as used in Figure~\ref{fig:adams}. In this plot we have used the modified versions of GSMFs of \hr,  after applying the SB observation limit together with the 30 pkpc aperture correction. }
\label{fig:Leja}
\end{figure}

In Figure \ref{fig:Leja} we compare the simulated GSMFs with more observations: \cite{Leja_2020} ({\it upper-left}), \cite{McLeod+21} ({\it lower-left}), \cite{Wright_2018} ({\it lower-right}), and \cite{Thorne+21} ({\it upper-right panel}). As stated above, the GSMFs derived by \cite{Leja_2020} have the highest amplitudes compared to the others. The GSMFs of \hr\ show that the size of the BCG population is comparable to the observations, except for that of \cite{Leja_2020}, whose BCG population size is substantially larger and, moreover, hardly shows any time evolution, which is in tension with the other empirical fits~\citep[for a detailed discussion, see][]{Leja_2020}.

\begin{figure}
\centering 
\includegraphics[width=\linewidth]{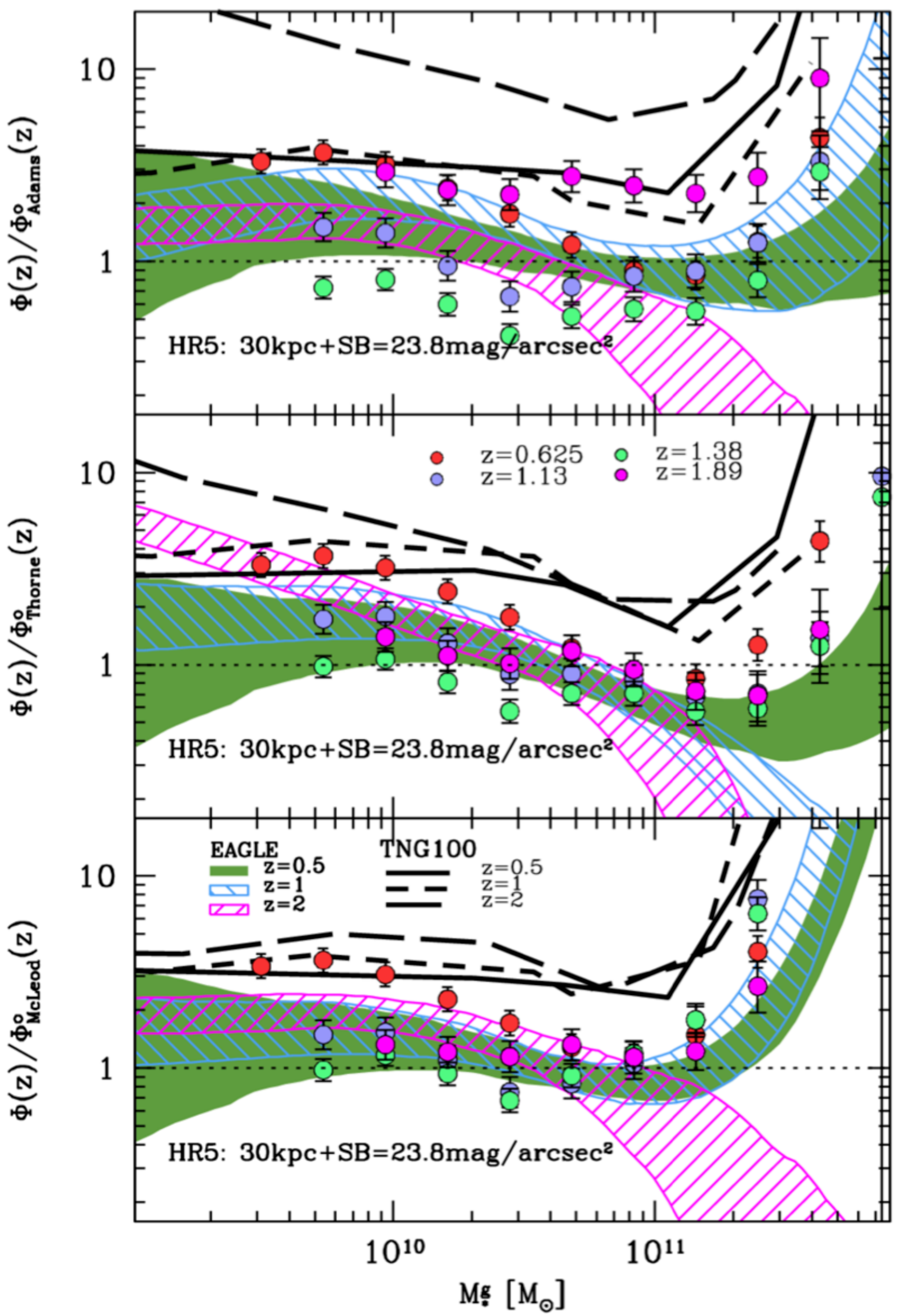}
\caption{Amplitude difference of the GSMFs between the \texttt{EAGLE}~\citep{furlong+15}, \texttt{TNG100}~\citep{pillepich18}, and \hr~\citep{lee21} simulations and observations of \cite{Adams21}, \cite{Thorne+21}, and \cite{McLeod+21} from the top to bottom, respectively. The base GSMFs of \hr\ are corrected with the 3D-aperture of 30 pkpc and the SB limit of 23.8 mag arcsec${}^{-2}$.}
\label{fig:mcleod}
\end{figure}

We measure the amplitude differences of GSMFs between observations and simulations. Figure~\ref{fig:mcleod} shows how much the GSMFs of simulations (marked by regions for \texttt{EAGLE}, lines for \texttt{TNG100}, and symbols with error bars for \hr)  deviate from the three different observations \citep{Adams21, Thorne+21, McLeod+21}. We note that the aperture correction and SB limit is applied only to \hr. Among the three simulations, \texttt{EAGLE} is least deviated from the three observations over the redshift range of $z\sim0.5-2$. The GSMFs of \hr\ shows considerable scatters around observations while the \texttt{TNG100} has a systematic overestimation on all mass scales. To reduce this systematic bias of \texttt{TNG100}, \cite{Tang+21} show that the systematic bias of \texttt{TNG100} can be significantly reduced by applying the SB limit and an apperture cut to the mock survyes in a similar way to ours.

\section{Discussions \& Conclusions} \label{sec:dis}

One of the key purposes of numerical simulations is to reproduce observations by tuning model parameters to understand underlying physics encoded in observed properties. This is  valid only when comprehensive understanding precedes for empirical data. %Thus, it is necessitated to evaluate the completeness of observed GSMFs in their mass ranges due to e.g. LSBGs~\citep[e.g.][]{wright+17}. 
In this study, we have found that the surface-brightness limit of galaxy surveys is one of the important factors fixing the survey completeness at the faint-end of GSMFs. For instance, with the SB limit of $\left<\mu_r\right>\le 25$ mag arcsec${}^{-2}$, our mock surveys reach about 65 \% completeness at the stellar mass of $M_\star^g=10^9 {\rm M_\odot}$ at $z=0.875$, while the completeness of the same mass drops sharply down to $\sim20\%$ at $z=1.89$. Even with $\left<\mu_r\right>\le 25$ mag arcsec${}^{-2}$, 80 \% of galaxies of $M_\star^g=10^9 {\rm M_\odot}$ are missed at $z=1.89$. One certainly needs stronger stellar and SN feedback models if observed GSMFs are simply accepted with no evaluation of the missing LSBGs at given SB limit. However, the fraction of LSBGs can be overestimated to some extent at the low-mass end due to the resolution limit of \hr.

%Nevertheless, if the observed GSMFs are found to be incomplete in the LSBG scale, numerical feedbacks should be rethought and simulation parameters need to be recalibrated.

Tantalizing evidences have been reported by a number of literature that are consistent with our claim of a substantial amount of missing LSBGs in galaxy surveys. \cite{greene+22} performed the image simulation to obtain 80 \% detection completeness at the SB limit of $\left<\mu_r\right>^e\sim $ 28 mag arcsec${}^{-2}$. %(here we assume the galaxy color of $(g-r)=0.5$). 
\cite{vanderburg+17} also showed that they reaches $\sim$80 \% completeness of LSBGs from ESO Kilo-Degree Survey with the SB limit of $\left<\mu_r\right>^e\sim$ 25.5 mag arcsec${}^{-2}$. For a lengthy list of complementary observational references, see also \cite{greene+22}. If the cosmological surface brightness dimming effect is added to the findings of \cite{vanderburg+17} for LSBGs at $z=0.625$ -- 2, neglecting the evolution correction, then it leads to the SB limit of about $\left<\mu_r\right>^e\sim $ 26 -- 28 mag arcsec${}^{-2}$. This is roughly consistent with the arguments presented in the main part of this paper. However, the global SFR may not change significantly after the SB limit corrections since the missing LSBGs insignificantly contribute to the global SFR in a low mass range~\citep{davies+16}. 

As of today, the ultra faint field at $\left<\mu_r\right> \gtrsim$ 28 mag arcsec${}^{-2}$ is an almost uncharted territory. The Dark Energy Camera (DECam) reaches $\left<\mu_r\right> = 28$ mag arcsec${}^{-2}$ in the Dark Energy Camera Legacy Survey  searching for LSBGs \citep{roman+21}. The Dragonfly Wide Field Survey also reaches 1$\sigma$ depths of $\left<\mu_g\right>\sim$ 31 mag arcsec${}^{-2}$ \citep{danieli+20}, and many extra-galactic survey projects (HST: \citealt{borlaff+19}; LSST: \citealt{brough+20}; Euclid: \citealt{euclid+22}) are on-going or scheduled for ultra faint-field observations of LSBGs. Recently, a space-born observatory, the JWST reaches the faint surface brightness limits of $\left<\mu_{\rm NIRCam}\right>= 31.1$--31.3 mag arcsec${}^{-2}$ at 3$\sigma$ fluctuations in $10{\arcsec}\times10{\arcsec}$ size with about 24-hour exposure \citep{montes+22}.

%[I FEEL THIS PARAGRAPH (WHICH I MAY BE WROTE), IS SLIGHTLY MISSPLACED. ANY SUGGESTIONS WHERE IT SHOULD GO? OR POSSIBLY REMOVED/SHORTENED?]
%Using the effective size, stellar mass and colors of the simulated galaxies, we calculated the cosmic surface dimming effect. We  matched the low-redshift evolution of the galaxy stellar mass functions on  dwarf galaxy scales. The overpopulation of brightest cluster galaxies in simulations could be mitigated by restricting  stellar mass within some distance to the galactic center. For redshift surveys  aimed at volume-limited sample up  to $z\simeq 1$ with the target minimum stellar mass $M_\star^g=10^{10} {\rm M_\odot}$, one should set the surface brightness limit of $\left<\mu_r\right>^e =$ 28 mag arcsec${}^{-2}$.  
Even with the correction to the LSB effects, the GSMFs of \hr\ still has a knee around the characteristic mass less distinct than that of observed GSMFs. It may come from an insufficient star formation efficiency ($\epsilon_\star=2\%$) or inefficient SN and AGN feedback. This calls for future studies based on the simulations with more complete physics and higher resolution to better understand the star formation and galaxy evolution on a wide-range of a mass scale.

Theoretical studies have suggested based on controlled hydrodynamical simulations that SFE ranges from 2\% to 9.5\%~\citep[e.g.,][]{matzner+02,semenov+16,kim+21}. We adopt $\epsilon_\star = 2\%$ in \hr, which is at the lower end of the SFE range. From a pair of simulations with two different SFE ($\epsilon_\star= 2\%$ \& 4\% ), we found that the simulation of $\epsilon_\star=4\%$ produces 1.5--3 times more galaxies than the simulation of $\epsilon_\star=2\%$ does at $z\simeq10$. The ratio however drops to $\sim1.3$ for $M\ge 10^9 ~{\rm M_\odot}$ at $z=1.8$ (see \ref{sec:sfecomp}). This indicates that higher SFE would not seriously change our fitting to $\mu_{\rm fit}^{\rm lim}$ at $z\lesssim5$, even though SFE is still a substantial factor regulating the growth of galaxies. As seen in Figure~\ref{fig:jwst}, \hr\ is also consistent with the current observations made by the JWST for the number density of massive proto galaxies at high $z$. Further observations and simulations are, however, required to statistically confirm the populations of massive proto galaxies in the current $\Lambda$CDM paradigm.

Also as shown in \ref{sec:sizediff}, the definition of the effective radius has a significant impact on the GSMF measurement in simulations. Since the three-dimensional effective radius is usually bigger than that measured in two dimensions, the galaxy SB is lower and, consequently, the GSMFs are more seriously affected by the SB limit. Likewise, the SB limit has more significant impact on the GSMFs when the effective radius is measured in the face-on plane, compared with that measured in the plane projected along the $x$ axis of the simulation box.

The simulation resolution, the seeing effect, and the source detection criteria are perhaps other important factors that influence the completeness of surveys. \hr\ is not able to resolve galaxies in the scales smaller than 1~pkpc. The resolution limit may severely affect early-type galaxies which have radii typically smaller than those of late types at given stellar mass~\citep[e.g.][]{nair10}. However, because of the relatively small fraction of early-type galaxies at $z\simeq1$~\citep[$\sim$15\%,][]{hwang+09}, the resolution limit of \hr\ would not significantly affect our results. The seeing effect is also able to influence the galaxy SB in observations (see Appendix~\ref{sec:seeing}). It reduces the observed SB of galaxies by more than 0.5 mag arcsec${}^{-2}$ when the effective radius of a galaxy is not sufficiently larger than the FWHM of the PSF. This may lead to the substantial uncertainty in deriving the SB limit in each survey catalog. Also, it is worthwhile to note that the pixel-level source detection limit may be another key factor in measuring the completeness of a galaxy catalog based on photometry (see Appendix~\ref{sec:source_detection}). If proper information on the source detection is provided, it would be more helpful in fine tuning the GSMFs. These three factors are closely related to each other, but to disentangle them is beyond the scope of this paper.

\acknowledgments

The authors thank the Korea Institute for Advanced Study for providing computing resources (KIAS Center for Advanced Computation Linux Cluster System) for this work. A special thank should be given to Profs. Ho Seong Hwang and Jihoon Kim who comment on the photometry of extragalactic images. JK was supported by a KIAS Individual Grant (KG039603) via the Center for Advanced Computation at Korea Institute for Advanced Study. JL is supported by the National Research Foundation of Korea (NRF-2021R1C1C2011626). This work benefited from the outstanding support provided by the KISTI National Supercomputing Center and its Nurion Supercomputer through the Grand Challenge Program (KSC-2018-CHA-0003). Large data transfer was supported by KREONET, which is managed and operated by KISTI. BKG acknowledges the support of STFC through the University of Hull Consolidated Grant ST/R000840/1, access to {\sc viper}, the University of Hull High Performance Computing Facility, and the European Union’s Horizon 2020 research and innovation programme (ChETEC-INFRA -- Project no. 101008324). This research was also partially supported by the ANR-19-CE31-0017 \href{http://www.secular-evolution.org}{http://www.secular-evolution.org}. YHK is supported by NRF-2022M3K3A1097100. High performance computing resources for this research were partially supported by the Research Solution Center and the National Supercomputing Center for Astrophysics and Space Sciences in the Institute for Basic Science.

\appendix
\section{Halo and Galaxy Findings}
\label{sec:finder}

\subsection{AMR \& Unified Datatypes}

Dark matter, super massive black holes (SMBHs), and stars are represented with point-mass  particles in \texttt{RAMSES}, while gas is a hydrodynamical component of a static structured mesh. Hence, in order to post-process the \ramses\ simulations, we integrate all the elements into a single unified data structure. We treat gas cells as particles that inherit all the mesh cell information. The center of a cell is set as the position of the cell `particle'. Although gas cells are hierarchically structured, we only consider leaf cells or terminal cells.

To reduce complexity in coding and enhance performance, we introduce a unified datatype for the four different matter species. The unified datatype has a fixed length of the common block containing the species, mass, position, and velocity, while extra space is arranged as a private block. The purpose of the private blocks is to save a union datatype for the raw simulation particle information, which has a variable length depending on the particle type (stellar, AGN, dark matter, and gas). The galaxy finding is processed with the data contained in the common block, while we save the finding results by dumping the data in the private blocks. 

\subsection{Adaptive Friend-of-Friend Method}

\begin{figure}
\centering 
\includegraphics[width=0.45\textwidth]{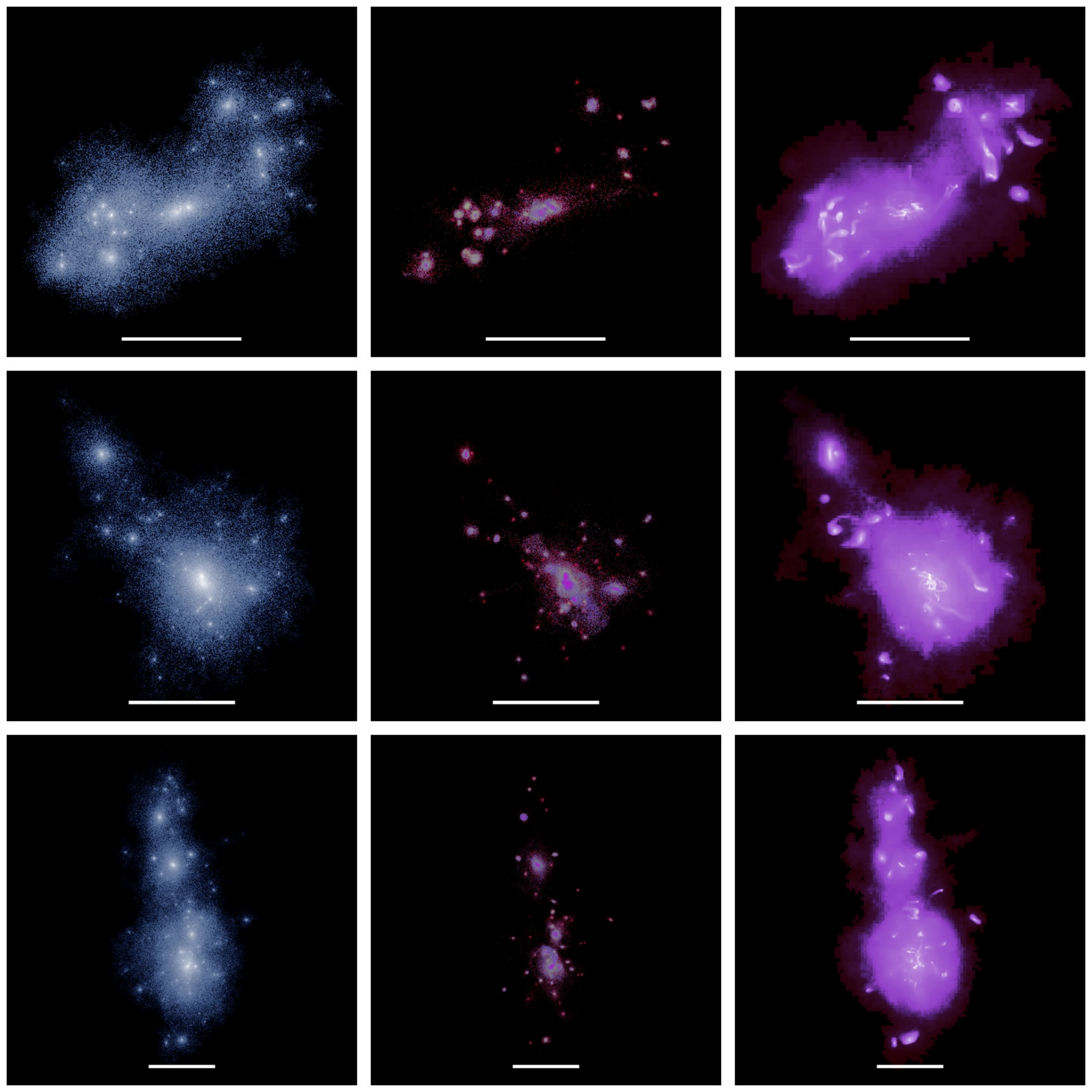}
\caption{Three projected density maps of massive halos identified with the variable length in the FoF at $z=0.625$. From the left column shown are the dark matter, stellar, and gas density fields.
In the middle panels, stellar particles with higher metallicity have  more pinkish color. The white bar at the bottom in each panel marks the scale of $l_{\rm bar}=0.5 ~h^{-1}{\rm cMpc}$. From the top-row panel, the total mass of the FoF halo is $1.09\times 10^{13} ~h^{-1}{\rm M_\odot}$, $1.42\times 10^{13}~h^{-1}{\rm M_\odot}$, and
$2.63\times 10^{13}~h^{-1}{\rm M_\odot}$, respectively.}
\label{fig:den_comp}
\end{figure}

To find virialized structures in cosmological $N$-body simulations, a percolation method like the Friend-of-Friend (FoF) algorithm is frequently employed. One fifth the mean particle separation is adopted as the standard linking length to identify structures from the distribution of uniform-mass particles. The linking length is known to find structures with a mean density of $\sim178$ times the critical density satisfying the cosmological virialization condition, according to the spherical top-hat collapse model \citep{suto+16}. The standard linking length is formulated as
\begin{equation} \label{eq:link}
	\ell_{\mathrm{link}} \equiv {1\over5} \left( { m_p \over \Omega_{m0} \rho_c }\right)^{1/3},
\end{equation}
where $m_p$ is the particle mass and $\rho_c$ is the critical density at the current epoch.

However, in $N$-body or hydrodynamical simulations, matter components may have varying mass not only between different types but also in a type. In this case, the standard linking length is needed to be modified to a general form. To identify FoF halos with varying mass particles, we use the averaged linking length between a pair of two particles of different types given as
\begin{equation}
	\ell_{\mathrm{comb}} = {1\over2}\left( { \ell_1 + \ell_2} \right),
\end{equation}
, where $\ell_1$ and $\ell_2$ are the linking lengths of the pair of each particle calculated by using Equation~\ref{eq:link}. This combined form is commutative or pair-wise mutual.

We apply this adaptive linking length to the \hr\ data of multiple particle types to identify FoF halos. Therefore, a FoF halo may consist of multiple types of matter components: dark matter, stellar, gas, and SMBH particles. Figure~\ref{fig:den_comp} show the images of three representative FoF halos identified at $z=0.625$. In this figure, a clear difference in density distributions are observed between the dark matter ({\it left}) and stellar components ({\it middle} column). The stellar distribution is more compact than the dark matter or gas. The tidal tails of gas ({\it right} column) in the merger remnants are most prominent among the tree components, and stellar streams between galaxies are also seen. The stellar density map is overlaid with the stellar metallicity colored in pink. From this, we can clearly observe that more massive galaxies are more metal enriched in the central regions. There is a good one-to-one correspondence among the density maps of the three different components, confirming that the adaptive linking length produces a consistent result from the distribution of different matter components.

\subsection{Halo Mass Function of \hr}
The halo mass function (HMF) is one of the fundamental statistics which closely relates to the cosmological models \citep{press74,veena+18,jenkins+01,Sheth+99}. It has been used as a fiducial statistic of $N$-body simulations, because it converges well between the $N$-body simulations of different resolutions at most redshifts. For example, Figure 7 of \cite{Kim+15} shows consistent fitting functions with less than a few percents of deviations.

\begin{figure}
\centering 
\includegraphics[width=0.45\textwidth]{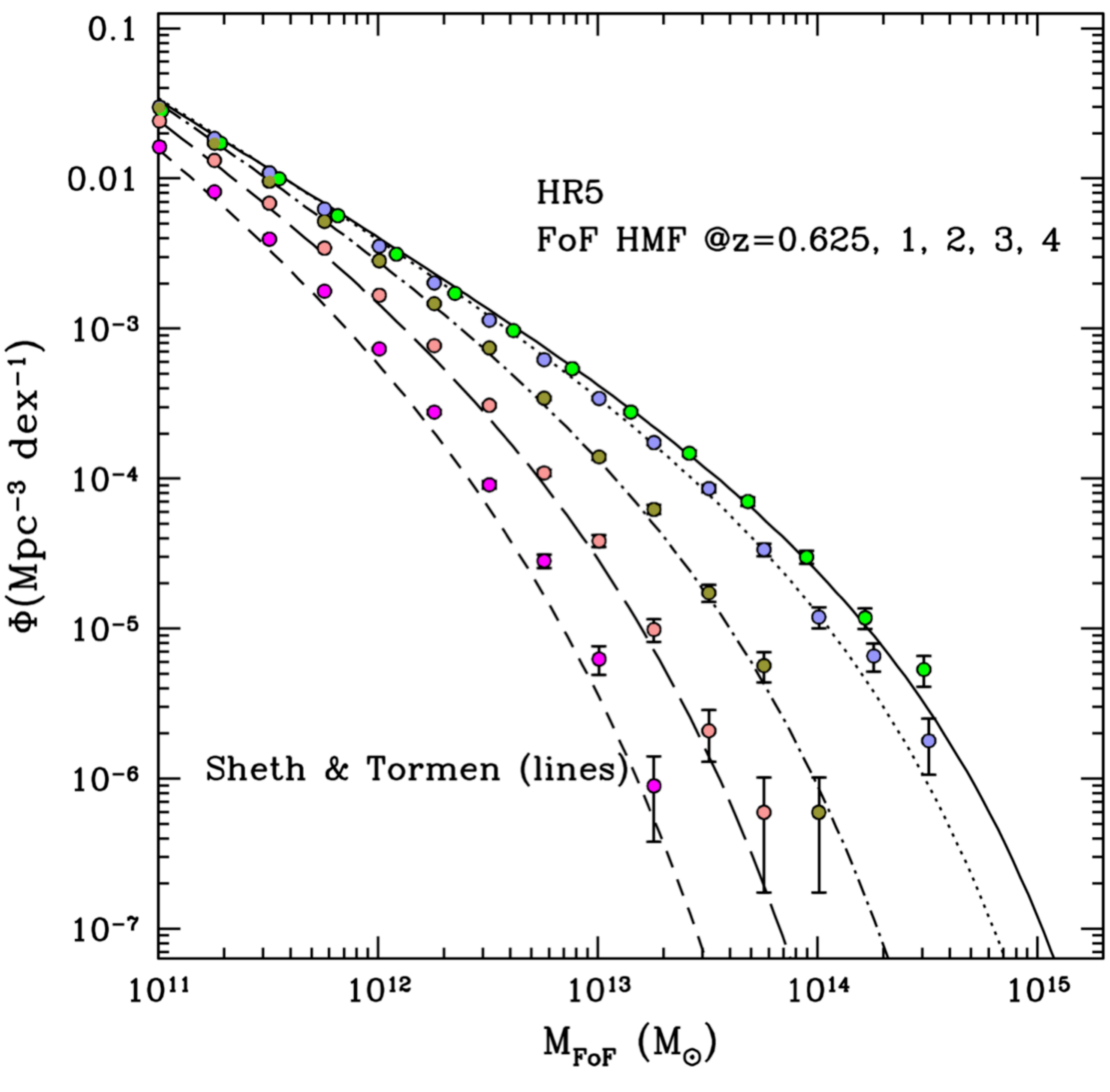}
\caption{Friend-of-Friend HMFs at several redshifts. Symbols with Poisson error bars are simulated HMFs while lines are the reference mass functions \citep{Sheth+99}.}
\label{fig:foflcdm}
\end{figure}

Throughout this paper, the mass function is defined as
\begin{equation}
	\Phi(M) \equiv  { \Delta N \over \Delta \log_{10} M}\,,
\end{equation}
where $\Delta N$ is the number of galaxies in a mass bin of size $\Delta \log_{10}M$.  In Figure \ref{fig:foflcdm}, we show the HMFs at $z=0.625$, 1, 2, 3, and 4. In the mass range of $10^{11}~{\rm M_\odot} \le M_{\rm FoF} \le 5\times 10^{14}~{\rm M_\odot}$, the \hr\ HMFs are well described by the reference model of the Sheth \& Tormen functions \citep{Sheth+99}.

\subsection{PGalF: A New Galaxy Finder}
We developed the \psb-based Galaxy Finder (\pgalf) based on the Physically Self-Bound (\psb) method, to identify galaxies composed of multiple matter components in \hr. We revisit \psb\ and describe the details of \pgalf\ in this subsection. 

\subsubsection{Revisiting the \psb\ Method}

\citet{kjhan06} first introduced \psb\ to identify self-bound structures (or subhalos) in their high-resolution $N$-body simulations~\citep{Kim+15,kjhan09}. \psb\ operates structure identification based on the coordinate-free density map and a web of network  made of neighbor links of particles. From the input number of neighbors, we build a network of neighboring. Densities are measured at all particle positions, with an adaptive smoothing kernel. This coordinate-free density field is advantageous over the regular density grid in several aspects. The result is independent of the size of the grids and it does not rely on a preferential shape of the particle distribution. 

On the coordinate-free density map, density peaks at particle positions are identified on the condition that all the neighboring particles should have lower density values. We define a core-density region around a peak by applying a water shedding method, to find the lowest threshold density for which no other density peaks are surrounded by the threshold contour. Then, each density peak may be extended to its core-density region, and we assume all particles in the region  become core members of the subhalo candidate. After extracting core particles, we apply the hierarchical density contours to separate the remaining particles into multiple hierarchical sets of particles. 

The aforementioned subhalo membership is mainly controlled by two conditions: the tidal boundary and total energy. If a particle is bound to a subhalo candidate and it is within the tidal boundary of the subahlo, it becomes the member of it. Unless it satisfies these two conditions, we then check the membership with respect to other subhalos. A particle may pass the membership check for two subhalos. Then, we set the particle to be a member of the less-massive subhalo.

\subsubsection{Tidal Radius}
There have been various definitions for the boundary in a binary system. \citep{binney+87, bosch18, renaud+16}. They depend on the assumptions made for the simplified model, such as the circularity of the orbital motion and point-mass or extended body of the host.

We parameterize the various definitions of the tidal radius to account for various situations. The tidal radius of a satellite can be generalized as \citep{kim06,bosch18} 
\begin{equation}
	r_t = R \left( { {m/M\over \alpha+\beta } }\right)^{1/3}, \label{eq:defrt}
\end{equation}
where $m$ is the satellite mass, $R$ is that distance from the host to the satellite, and $M$ is the host mass contained within $R$. We introduce a dimensionless parameter, $\alpha$,  to account for the mass distribution of the host at the satellite position as
\begin{equation}
	\alpha(R) \equiv 2 - \left({d\ln M\over d\ln R}\right)\bigg\rvert_R.
\end{equation}
Therefore, if the host is a point mass,  we get $\alpha=2$.
In Equation~\ref{eq:defrt}, $\beta$ is a function of orbital motion reflecting the effect of centrifugal force on the tidal radius, and can be formulated as \citep{bosch18},
\begin{equation}
	\beta \equiv {\Omega^2 R^3 \over GM},
\end{equation}
where $\Omega (\equiv V_c/R)$ is the angular velocity \citep{king62} for the circular velocity, $V_c$. From this equation, one may easily derive that $\beta = -2\mathcal{K}_t/\mathcal{W}$ where $\mathcal{K}_t$ is the rotational kinetic energy of the satellite and $\mathcal{W}$ is its potential energy. For a circular orbit, $\beta=1$ but $\beta=0$ for a radial motion. If the satellite is not bound to the host, then we neglect this effect and set $\beta=0$. From this parameterization we recover nearly all the modeled radii of the satellite. For example, we recover the Jacobi radius with $\alpha+\beta=3$ for a satellite in a circular orbit around a non-contacting host,  or we recover the Roche limit (when $\alpha+\beta=2$) for a radial orbit around the host.

Now, we discuss the $\alpha$ parameter in more detail for the cosmological extended object. Assuming the NFW density profile \citep{Navarro+97} for a host with virial mass, $M_v$, the enclosed mass, $M(s)$, is modeled as \citep{lokas01},
\begin{equation}
	M(s) = M_v g(c) \left[ \ln (1+cs) - {cs\over 1+cs} \right], 
\end{equation}
where $c$ is the concentration index, $s$ is the scaled radius to the virial radius as $s\equiv R/R_v$, and
\begin{equation}
	g(c) \equiv \left[ \ln(1+c) - c/(1+c)\right]^{-1}.
\end{equation}
Then, we finally get
\begin{equation}
	\alpha (s) \equiv 2  - \left( { cs\over 1+cs}\right)^2.
\end{equation}
The concentration index is empirically given as \citep{comerford07}
\begin{equation}
	c = 14.5 \left({M_v\over 1.3\times 10^{13}~h^{-1}{\rm M_\odot}}\right)^{-0.15}(1+z)^{-1} \,,
\end{equation}
where $z$ is the redshift and $h$ is the Hubble expansion rate divided by 100 km/second/Mpc.

\subsubsection{Hierarchical Membership Determination}
\begin{figure}
\centering 
\includegraphics[width=0.45\textwidth]{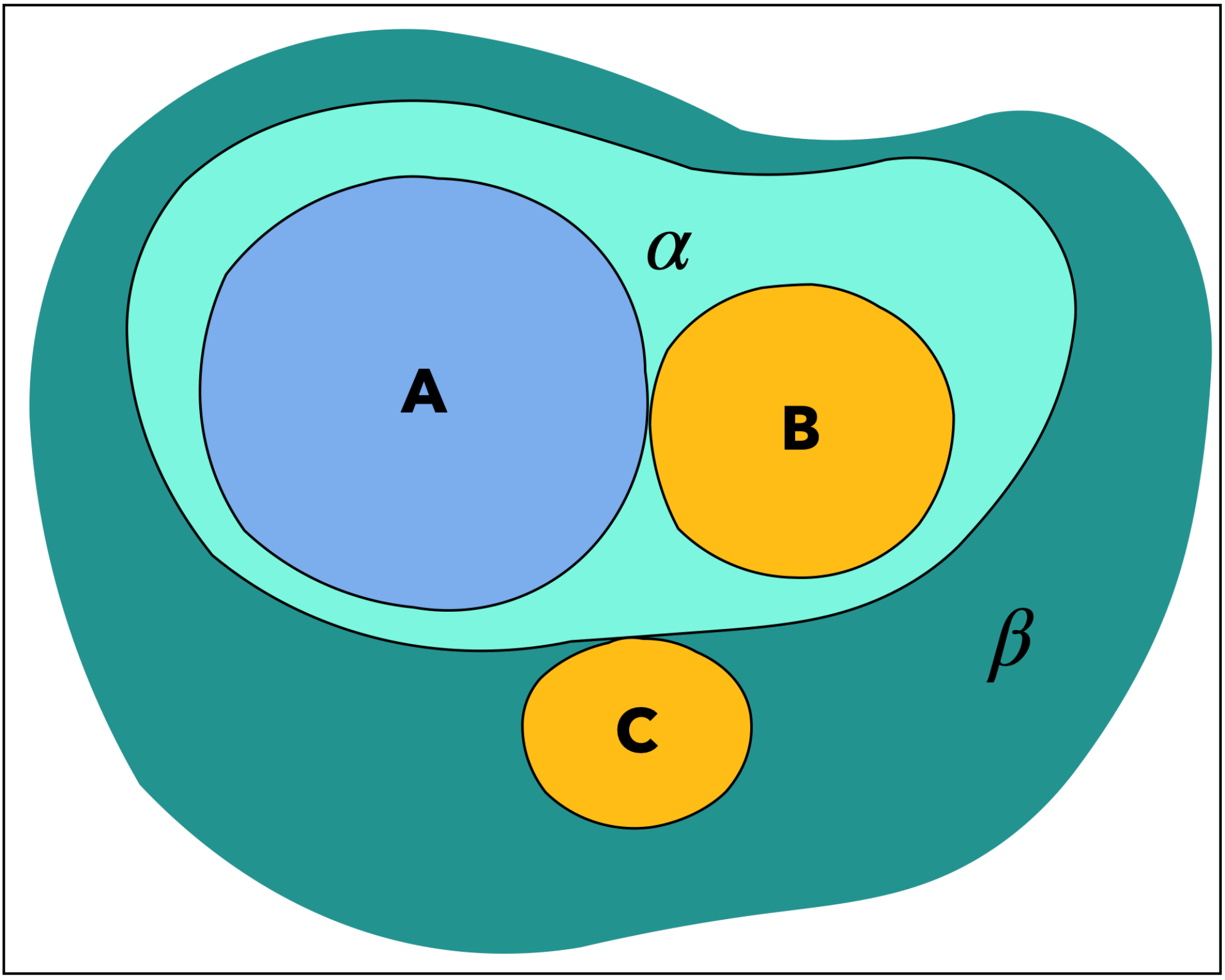}
\caption{Two-dimensional example of the density field and iso-density contours which split and arrange areas in the hierarchical manner. The density field is made using a virialized FoF halos using its particles.}
\label{fig:pbs_density}
\end{figure}

Figure \ref{fig:pbs_density} exemplifies how the subhalo finding of \psb\ is working for a given density field. In this density map, there are three density peaks and each peak has the corresponding core region marked with {\bf A, B}, and {\bf C}. First, the {\bf A} and {\bf B} regions are found separated by the inner most contour while the {\bf C} area is delineated by another contour that separates the {\bf C} region from the other two core regions. This separation is done via a water-shedding technique. It is a percolation method which grows a volume of interest by lowering the density threshold around the target density peak. Therefore, the regions {\bf A} and {\bf B} are specified by the density threshold given by the saddle point between those nearby density peaks. Particles in the core regions are named core particles and they are assumed to be members of  the subhalo candidate. From now on, we mix the use of {\bf A, B} and {\bf C}, to denote core regions or corresponding subhalo candidates depending on the meaning in the text. 

Particles in the shell region of $\boldsymbol{\alpha}$ are supposed to be members of one of two halo candidates, {\bf A} and {\bf B}. We calculate the tidal radius of the less massive subhalo and check whether a particle in the $\boldsymbol{\alpha}$ shell region lies within the tidal radius and its total energy with respect to the subhalo is negative. If not, we move on to the next more massive subhalo and judge its membership in the same manner. At last, particles in the $\boldsymbol{\beta}$ region are picked up to check  membership with respect to the subhalos, {\bf A, B}, and {\bf C}.

Whenever we move our focus to the next shell particles, we update the tidal radius of subhalos and re-apply the membership determination for  member particles to confirm whether they could still hold their membership under the  updated circumstances.

\subsubsection{Stellar Density Field in \pgalf}
Like the \psb\ method, the \pgalf\ is based on the coordinate-free density map and a network of particle neighbors. In \pgalf, however, the galaxy finding is based on the stellar mass density, instead of the dark-matter mass density used in the analysis for pure $N$-body simulations. The density kernel, W4, is also used for a fixed number of nearest stellar particles. However, a neighboring link can be made between different matter species in \pgalf. Except for this, the overall scheme is similar to the \psb\ method.

\begin{figure}
\centering 
\includegraphics[width=0.45\textwidth]{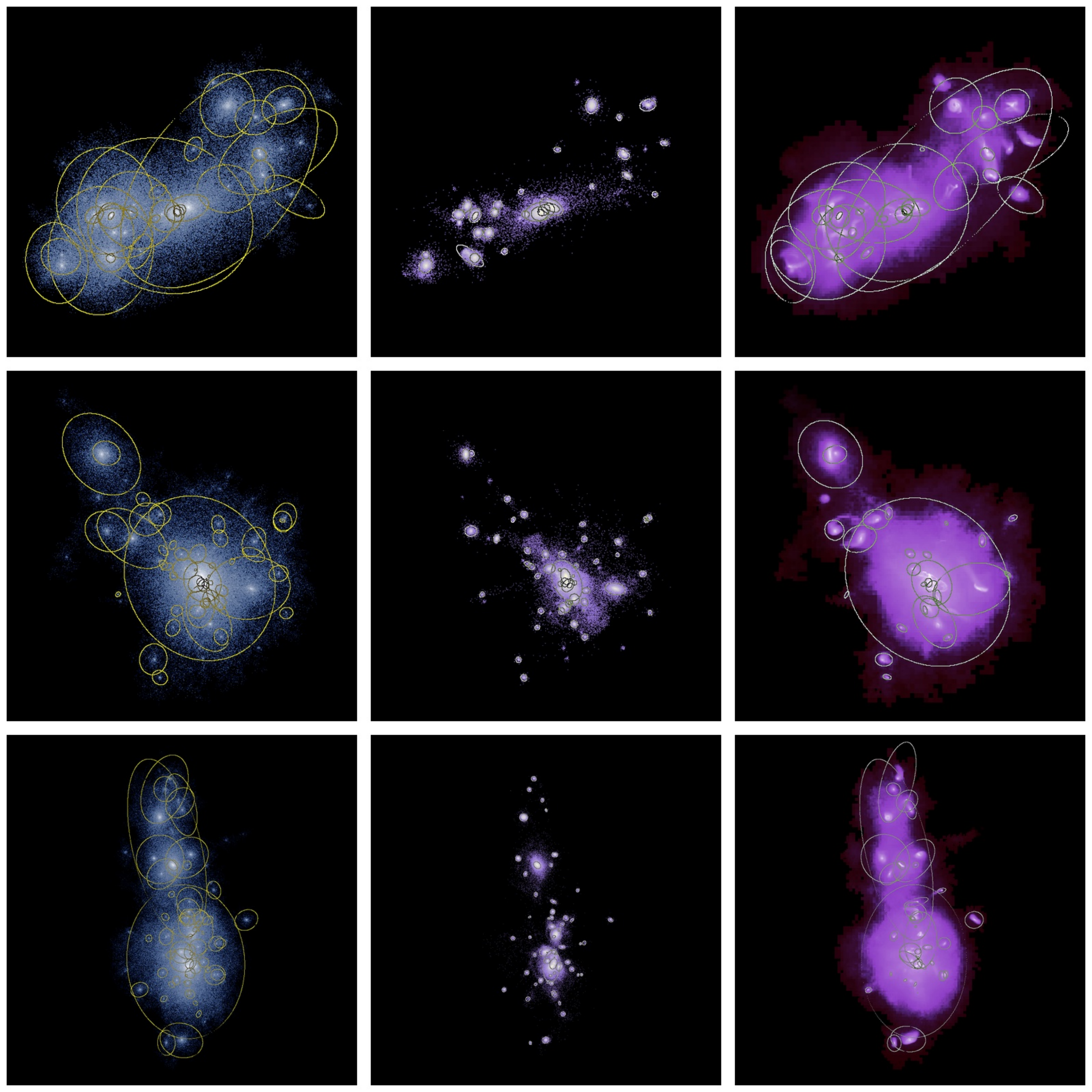}
\caption{Galaxy finding results applied to the FoF halos as given in Figure~\ref{fig:den_comp}. Contours are shown after fitting the density distribution of each component of galaxies.}
\label{fig:gal}
\end{figure}

The stellar density-based approach has several advantages. First, stellar distributions are more compact than those of dark matter, which helps us clearly detect the boundary of a galaxy system. Also, stellar components of galaxies tend to be at the near bottom of potential well, and this helps to easily add  member particles starting from the core region. Also the substantial parts of the stellar component are not bound to galaxies (see the panels in the middle column of Figure~\ref{fig:gal}) due to the dynamical frictions and tidal stripping in a group or cluster region.

\section{Difference between the absolute magnitudes in the SDSS $r$- and Cousin $R$-bands}
\label{sec:diffmag}

\begin{figure}
\centering 
\includegraphics[width=0.45\textwidth]{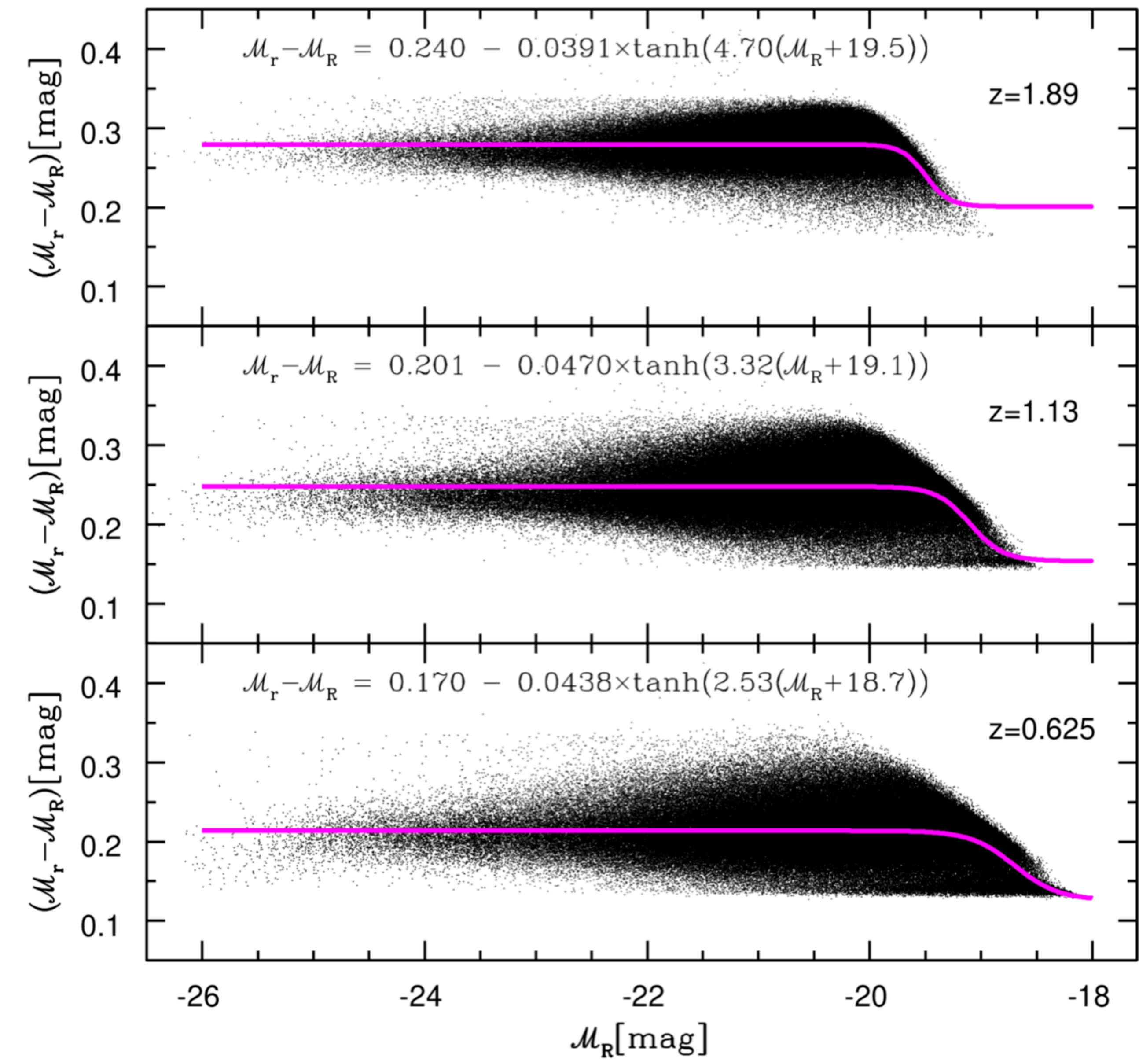}
\caption{Difference between the absolute magnitudes in the SDSS $r-$ and Cousin $R-$bands. Magenta solid lines present the fitting functions derived from the $\mathcal{M}_r-\mathcal{M}_R$ distribution using the $\chi^2$ minimization.}
\label{fig:r_R}
\end{figure}

We have examined the difference between the absolute magnitudes of the \hr\ galaxies of $M_{\star}>10^9\,\msun$ in the SDSS $r-$ and Cousin $R-$bands at $z=0.625$, 1.13, and 1.89 (see Figure~\ref{fig:r_R}). The $R$-band absolute magnitude $\mathcal{M}_{R}$ is slightly lower than $\mathcal{M}_r$ and the difference becomes smaller with decreasing redshifts. We present the $\chi^2$-minimization fitting result with a magenta solid curve in each panel.

\section{Observational Constraints on the Source Detection}
\label{sec:source}
\subsection{Pixel-Scale Source Detection Criteria}
\label{sec:source_detection}

The source detection in the photometry of extragalactic images usually begins with searching image pixels that have values larger than $f_{\rm p} \sigma_{\rm p}$ where $f_{\rm p}$ is the multiplication factor above the background noises, ($\sigma_{\rm p}$). The pixels are grouped by a given connection criterion, being regarded as a galaxy candidate. Then, an aperture with the solid angle $A_{\rm ap}$ is located at the center of the candidate to calculate the total flux by integrating the pixel values enclosed in the aperture. Usually, there is a criterion to limit the faint-end magnitude ($m_{\rm ap}^{\rm lim}$) of galaxy candidates. The average pixel value should be $f_{\rm ap}$ times larger than the background fluctuations on the aperture scale ($\sigma_{\rm ap}$). For the Gaussian fluctuations of background, we may assume the scaling relation between these two noises as, $\sigma_{\rm p} = \sigma_{\rm ap} (A_{\rm ap}/A_{\rm p})^{1/2}$, where $A_{\rm p}$ is the solid angle of the pixel. Finally we may formulate the SB limit of the source detection as \citep[also see][]{roman+20},
\begin{equation}
    \mu_{\rm p}^{\rm lim} = -2.5 \log\left[{ {f_{\rm p}\over f_{\rm ap}} 
    \left({A_{\rm ap} \over A_{\rm p}}\right)^{1/2}  }\right] + \mu_{\rm ap}^{\rm lim}
\end{equation}
where the average SB of the aperture is related to the galaxy magnitude limit as $\mu_{\rm ap}^{\rm lim} \equiv m_{\rm ap}^{\rm lim} + 2.5\log(A_{\rm ap}/{\rm arcsec^2})$. 

\begin{figure}
\centering 
\includegraphics[width=0.45\textwidth]{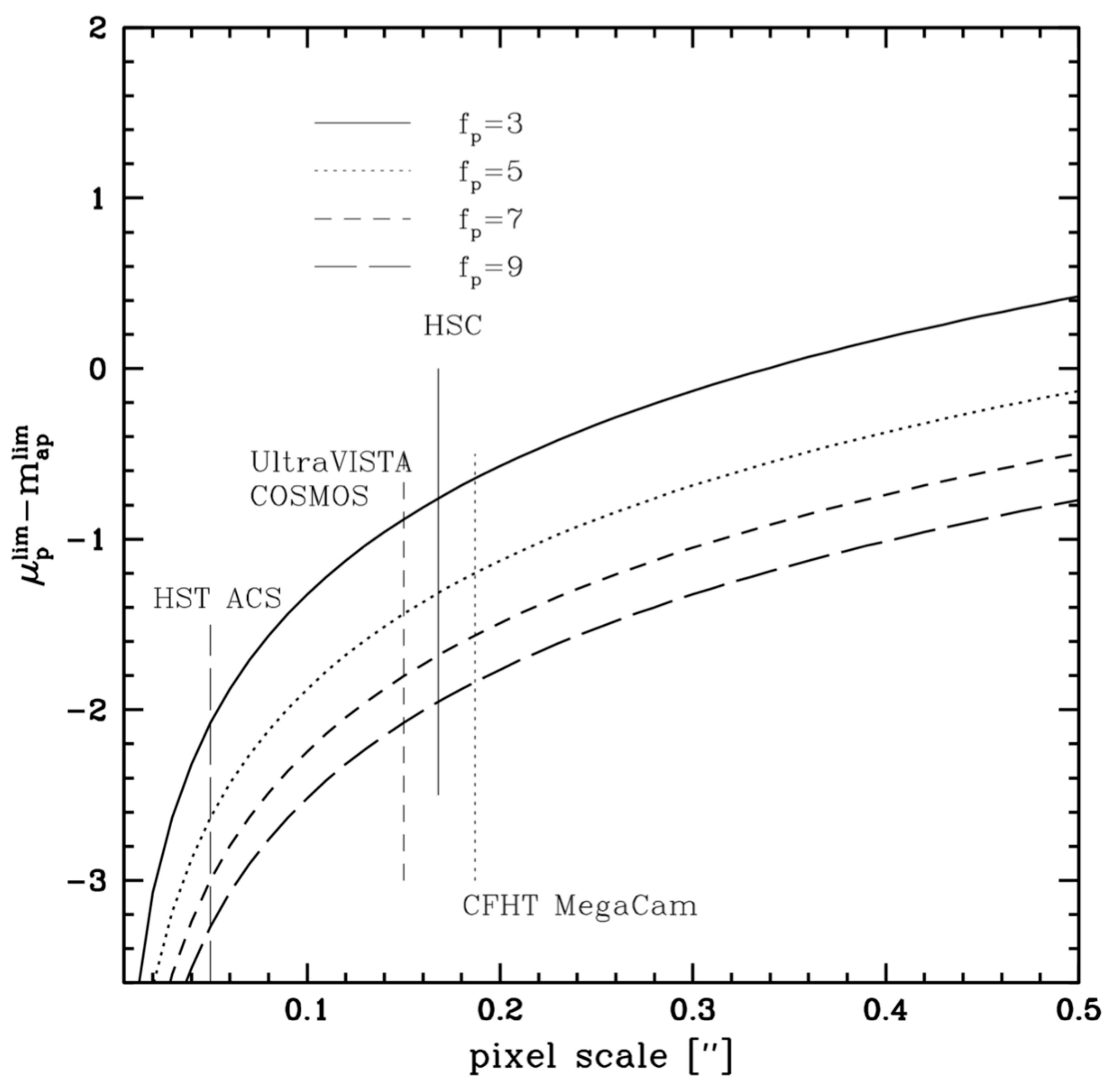}
\caption{The derived SB limit as a function of the pixel scale for several $f_{\rm p}$. The solid, dotted, short-dashed, and long-dashed vertical lines mark the pixel scales of Hyper Suprime-Cam \citep{aihara+18...70S...4A}, CFHT MegaCam, UltraVISTA/COSMOS \citep{mccracken+12}, and Advanced Camera for Survey (ACS) of HST, respectively. }
\label{fig:sb}
\end{figure}

Figure~\ref{fig:sb} shows the offset between $\mu_{\rm p}^{\rm lim}$ and $m_{\rm ap}^{\rm lim}$ as a function of the pixel scale with the aperture size ($r_{\rm ap}=1{\arcsec}$) and the source-detection level ($f_{\rm ap}=5$), which are recommended by \texttt{SExtractor} \citep{bertin+96}. For the source detection criterion of $3\sigma_{\rm p} \le \mu_{\rm p}^{\rm lim}\le 5\sigma_{\rm p}$, the SB limit would be about 1-- 2 magnitude brighter than the aperture magnitude limit for the faint galaxy photometry of the \texttt{COSMOS} field.

\subsection{Seeing Effects}
\label{sec:seeing}

\begin{figure}
\centering 
\includegraphics[width=0.45\textwidth]{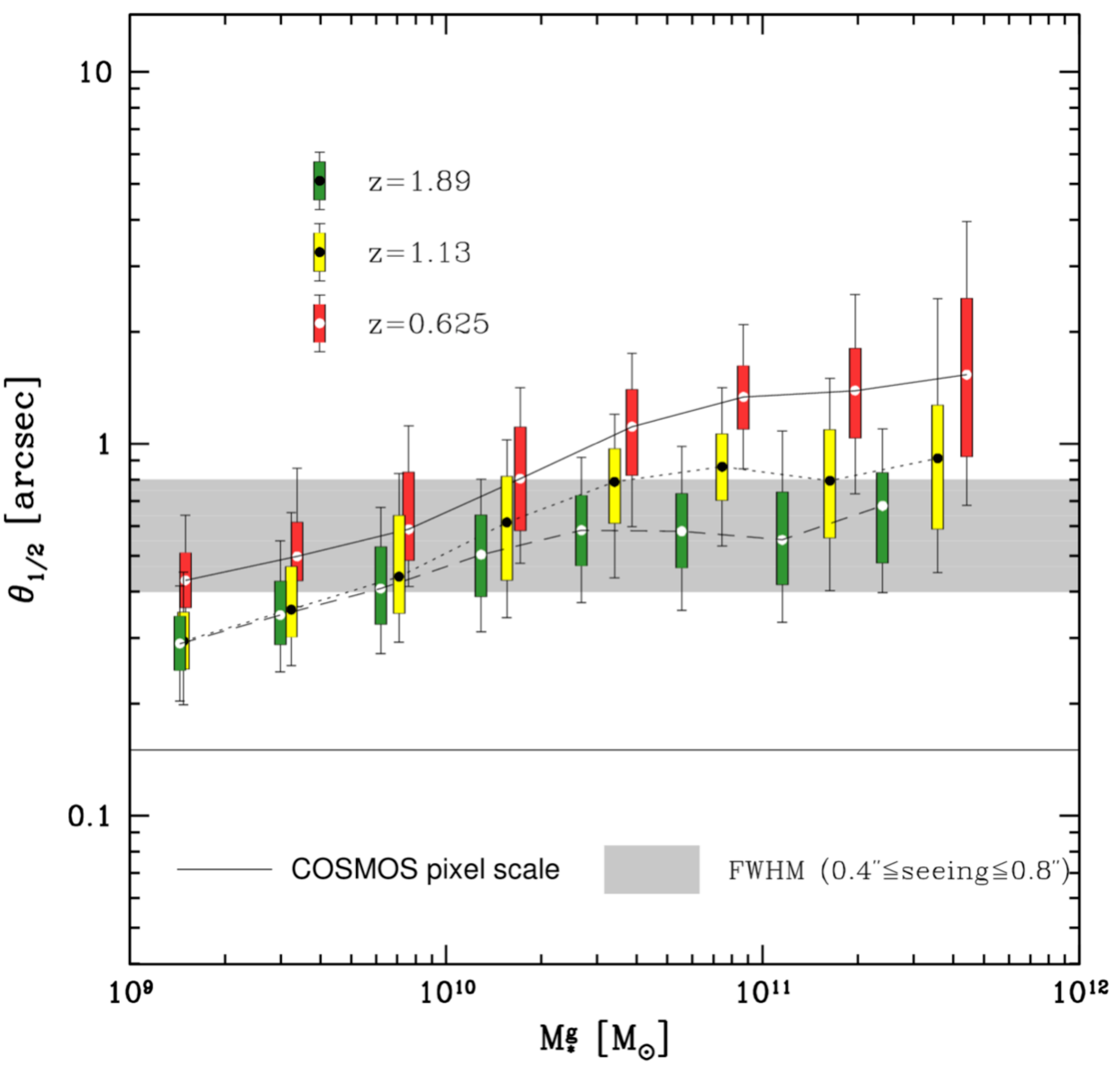}
\caption{The angular scales of simulated galaxies (candle bars), pixel (horizontal bar), and average seeing at low-$z$. The gray region delineates the distribution of FWHM of PSF in the UltraVISTA observations \citep{mccracken+12}.}
\label{fig:seeing}
\end{figure}

In this subsection, we briefly discuss the seeing effects on the galaxy SB. First, in Figure \ref{fig:seeing}, we compare the angular scales of simulated galaxies (candle bars), COSMOS pixel ($0{\arcsec}\!\!.15$; a solid horizontal bar), and the range of the observed FWHM of a PSF ($0{\arcsec}\!\!.4$--$0{\arcsec}\!\!.8$; a gray region). Even though the pixel scale of, for example, the COSMOS field is sufficiently smaller than the galaxy angular size, the galaxy shape and SB are severely contaminated by the seeing effect~\citep{mccracken+12}. 

By simulating the seeing effect on the galaxy models with S\'ersic profiles, \cite{trujillo+01} showed that the observed SB becomes fainter than the input value by $\Delta \mu^{\rm seeing} \gtrsim 0.5$ when the galaxy angular size is comparable to or smaller than the FWHM of the PSF.

\section{Effect of Star Formation Efficiency on the GSMF} \label{sec:sfecomp}
In this section, we examine the effect of star formation efficiency $\epsilon_\star$ on the GSMF using two auxiliary simulations. The first simulation (\texttt{M2}) shares the same simulation parameters with \hr\ except for the smaller box size, $L^{\rm aux}_{\rm box}=128$ cMpc. The second simulation (\texttt{M4}) is intended to study the role of SFE having the same parameters as of \texttt{M2} except for $\epsilon_\star$= 4 \% without any significant systematic bias. Both simulations are run down to $z=1.7$.

\begin{figure}
\centering 
\includegraphics[width=0.45\textwidth]{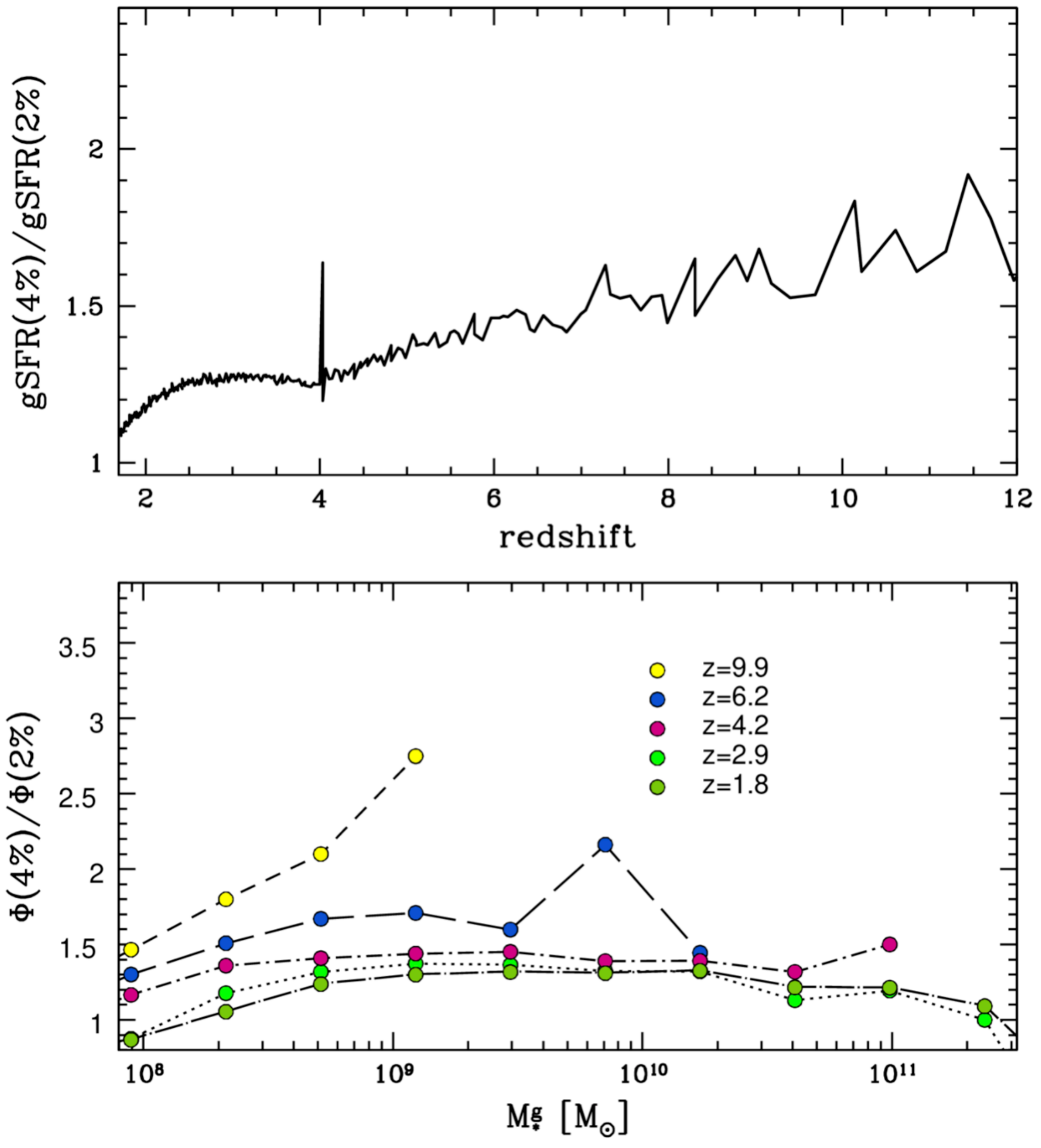}
\caption{The ratios of the global SFR (top) and GSMFs (bottom panel) between \texttt{M2} and \texttt{M4}. The spurious spike at $z=4$ is caused by the discrete global mesh refinement scheme of \ramses\, in the global SFR ratio.}
\label{fig:offset}
\end{figure}

The global SFR is directly regulated by $\epsilon_\star$, producing, as a result, different GSMFs. The top panel of Figure \ref{fig:offset} shows the ratio of the global SFRs between \texttt{M2} and \texttt{M4} as a function of redshift. The ratio decreases over time, but it is still higher than unity at $z\sim2$. We also compare the GSMFs of the two simulations in the bottom panel. The difference gradually decreases with decreasing redshift at least until $z=1.8$. This is because higher SFE essentially accompanies more energetic stellar feedback which blows gas from galaxies while baryon reservoirs are the same between the two. Note that the volume of the two simulations is not large enough to make the statistic of GSMFs at $z>10$.

\section{Dependence of GSMFs on the Effective Radius}\label{sec:sizediff}
\begin{figure}
\centering 
\includegraphics[width=0.45\textwidth]{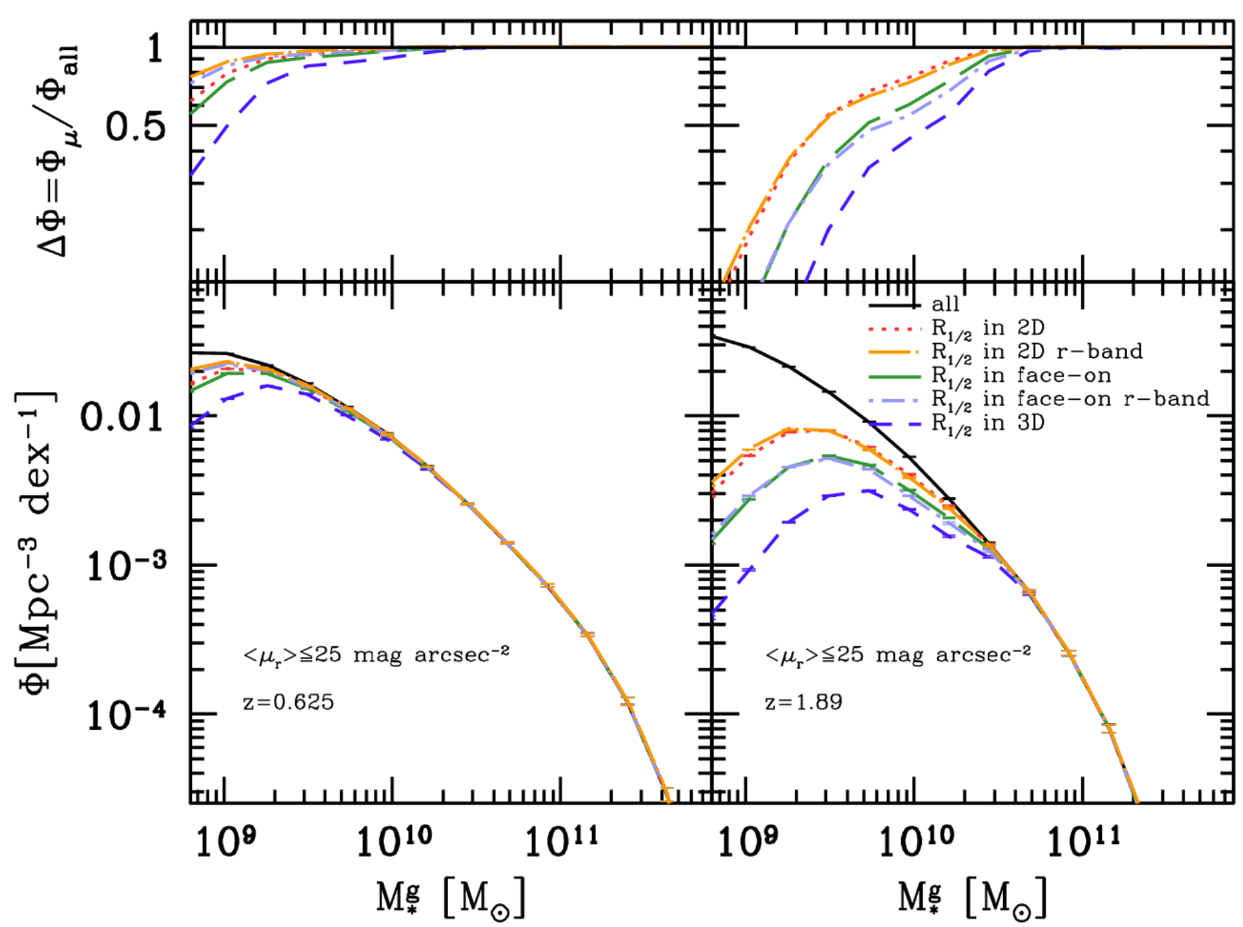}
\caption{The simulated GSMFs for different definitions of the galaxy radius. On the bottom panel shown are GSMFs for the cases of effective radius measured in projected distributions of mass (dotted) and $r$-band brightness (long dot-dash). On the other hand, the GSMFs with the radius measured in the face-on view of a galaxy is shown in long-dashed (mass) and short dot-dashed ($r$-band) lines. The short-dashed line is for the case of radius when measured in three dimension. The true GSMF is shown in the thick solid line (all). On the top panel, we show the difference of GSMFs with respect to the true GSMF. In this plot, we use the SB cut of $\mu_r^{\rm lim}=25$ mag arcsec${}^{-2}$.}
\label{fig:sizediff}
\end{figure}

The effective radius of a galaxy can be defined in various ways, and thus we examine in this section how the different definition may influence the resulting GSMFs. Figure \ref{fig:sizediff} shows the GSMFs derived based on five different definitions of the effective radius at $z=0.625$ (left column) and $z=1.89$ (right column). Because the effective radius is generally larger when it is measured in three dimensions than in two dimensions, the GSMFs are most significantly lowered with $R_{1/2}$ in 3D (dashed purple). Likewise, the GSMFs are diminished more due to the SB limit in the face-on projection than in the case of projection along x-axis which effectively random projection. On the other hand, there is no substantial difference between the radii measured from stellar mass and $r$-band luminosity distribution.

\end{document}